\documentclass[twocolumn,twocolappendix]{aastex631} 
\usepackage[utf8]{inputenc}
\usepackage{enumerate}
\usepackage{graphicx}
\usepackage{hyperref}
\usepackage{mathtools}
\usepackage{comment}
\usepackage{tabularx}
\usepackage{amsmath}
\usepackage{changepage}
\usepackage{ amssymb }
\usepackage{lipsum} 
\usepackage{float}
\usepackage{commath}
\usepackage{enumerate}
\usepackage{longtable}
\usepackage{stackengine}

\usepackage{cleveref}
\usepackage[caption=false]{subfig}
\usepackage{placeins}
\usepackage[ruled,lined]{algorithm2e}
\newcommand{\sgn}{\operatorname{sgn}}
\AfterEndEnvironment{figure}{\noindent\ignorespaces}

\usepackage{xcolor}

\begin{document}
\title{Constraining the Gravitational Potential \\ from the Projected Morphology of Extragalactic Tidal Streams} 
\begin{abstract}
The positions and velocities of stellar streams have been used to constrain the mass and shape of the Milky Way’s dark matter halo. Several extragalactic streams have already been detected, though it has remained unclear what can be inferred about the gravitational potential from only 2D photometric data of a stream. We present a fast method to infer halo shapes from the curvature of 2D projected stream tracks. We show that the stream’s curvature vector must point within 90 deg of the projected acceleration vector, in the absence of recent time-dependent perturbations. While insensitive to the total magnitude of the acceleration, and therefore the total mass, applying this constraint along a stream can determine halo shape parameters and place limits on disk-to-halo mass ratios. The most informative streams are those with sharp turns or flat segments, since these streams sample a wide range of curvature vectors over a small area (sharp turns) or have a vanishing projected acceleration component (flat segments). We apply our method to low surface brightness imaging of NGC 5907, and find that its dark matter halo is oblate. Our analytic approach is significantly faster than other stream modeling techniques, and indicates what parts of a stream contribute to constraints on the potential.  The method enables a measurement of dark matter halo shapes for thousands of systems using stellar stream detections expected from upcoming facilities like Rubin and Roman.
\vspace{1cm}
\end{abstract}

\author[0000-0001-8042-5794]{Jacob Nibauer}\email{jnibauer@princeton.edu}
\affiliation{Department of Astrophysical Sciences, Princeton University, Princeton, NJ 08544, USA}
\affiliation{The Observatories of the Carnegie Institution for Science, 813 Santa Barbara Street, Pasadena, CA 91101, USA}

\author[0000-0002-7846-9787]{Ana Bonaca}
\affiliation{The Observatories of the Carnegie Institution for Science, 813 Santa Barbara Street, Pasadena, CA 91101, USA}

\author[0000-0001-6244-6727]{Kathryn V. Johnston}
\affiliation{Department of Astronomy, Columbia University, 550 West 120th Street, New York, NY 10027, USA}
\affiliation{Center for Computational Astrophysics, Flatiron Institute, 162 5th Av., New York City, NY 10010, USA}

\section{Introduction}\label{sec:intro}
Low surface brightness imaging has revealed intricate tidal features in the halos of external galaxies (e.g., \citealt{1998ApJ...504L..23S,2001Natur.412...49I,2009ApJ...692..955M,2016ApJ...823...19C, 2018A&A...614A.143M, 2018ApJ...866..103K,2023arXiv230204471G, 2023A&A...671A.141M}). Similar tidal features have been precisely measured in the Milky Way, where the field of tidal debris can be studied on a star-by-star basis (e.g., \citealt{2003ApJ...599.1082M,2003AJ....126.2385O,2006ApJ...642L.137B, 2010ApJ...712..260K,2019ApJ...872...58H,2022ApJ...926..107M,2011MNRAS.417..198V,2022arXiv221104495K}). The morphology of these systems is vast, ranging from streams, to shells, and continuous deformations of the two. These features originate from tidal stripping, when a lower mass satellite galaxy accretes onto the more massive host. 


The phase-space distribution (that is, the space of positions and velocities) of tidal debris is sensitive to the underlying gravitational potential of the host galaxy. Stellar streams, for instance, roughly trace orbits in the host gravitational potential, and provide a sensitive probe of both the baryonic and dark matter components of a galaxy. An extensive range of methods have been devised to measure the gravitational potential using tidal debris. These range from trial orbit integrations (e.g., \citealt{1999ApJ...512L.109J,2006MNRAS.366.1012F,2010ApJ...712..260K,2014ApJ...794....4P}) to fully generative models for stream formation (e.g., \citealt{2012MNRAS.420.2700K,2015MNRAS.452..301F,2014ApJ...795...94B,2014MNRAS.445.3788G}). In the Milky Way some of these methods have been applied to real stellar streams, where kinematic information and distance tracks can be utilized to place tight constraints on the potential (e.g., \citealt{2019MNRAS.486.2995M, 2022arXiv221104495K}).

From an information theoretic perspective, \cite{2018ApJ...867..101B} found that kinematically cold streams provide a localized constraint on the galactic potential. Indeed, in \cite{Nibauer} it is shown that local galactic accelerations can be recovered from streams directly, without explicit reference to any potential model. The results from these analyses both imply that there must be some amount of information about the potential encoded in the geometry of tidal features (e.g., the light blue contours in Fig.~3 of \citealt{2018ApJ...867..101B} and Eq.~6 of \citealt{Nibauer}), though it has remained unclear what properties of the potential can actually be inferred from projected stream morphology. Using the generative ``streakline" technique, \citet{2015ApJ...799...28P} found that a spherical halo reproduces the curvature of Palomar 5 better than a triaxial halo. This method requires one to model the full time-evolution of the progenitor, introducing significant degeneracies with initial conditions, mass-loss rates, and integration times. If the object of interest is the shape of the gravitational potential, avoiding these degeneracies would be optimal. Especially in the context of external galaxies, attempting to model the full time-evolution of a progenitor from only the projected shape of a tidal feature introduces far more modeling degrees of freedom than the data can realistically constrain.

For the thousands of extragalactic tidal features expected to be discovered with upcoming surveys like The Rubin Observatory \citep{2019ApJ...873..111I}, {\it Euclid} \citep{2011arXiv1110.3193L}, and {\it The Nancy Grace Roman Space Telescope} ({\it Roman}; \citealt{2013arXiv1305.5422S}), neither distance tracks nor velocity information will be available for most systems. In this limited data regime, modeling attempts will be hindered by degeneracies with projection effects and the unobserved phase-space dimensions. For instance, action-angle based methods also require measurements of the 6D phase-space distribution of the tidal feature \citep{2013MNRAS.433.1826S, 2015ApJ...801...98S}, severely limiting applicability to external galaxies. While a single extragalactic stream will not be measured to the same level of detail as in the Milky Way, measurements of the low surface brightness universe provide a unique opportunity to constrain dark matter halo properties at the population level. In this context, automated methods will be required to characterize the ensemble of tidal debris and translate observations to physical constraints on the dark matter distribution in external galaxies. 

 Recently, \cite{2022ApJ...941...19P} modeled the tidal feature surrounding the galaxy Centaurus A using the ``particle spray" method \citep{2015MNRAS.452..301F}. This method provides a semi-analytic prescription for stream-formation, and works by releasing test particles from the Lagrange points of a progenitor throughout a trial orbit integration. In the limited-data scenario of an external galaxy, \cite{2022ApJ...941...19P} demonstrates the significant degeneracies in modeling the projected morphology of an extragalactic stream without velocity information. Indeed, the shape of a stream is degenerate with mass enclosed, halo shape parameters, velocity, and projection effects. \cite{2022ApJ...941...19P} showed that if a single radial velocity is measured along the stream, some of these degeneracies (the halo mass, in particular) can be broken.  
 
Other attempts to model the potential of external galaxies using extragalactic tidal features have relied on full $N-$body models for stream (or shell) formation \citep{2013MNRAS.434.2779F,2014MNRAS.442.3544F,2022A&A...660A..28B,2023A&A...669A.103M}, additional applications of the ``particle spray" technique \citep{2015arXiv150403697A,2019ApJ...883L..32V}. In total, generative stream models have a large number of latent nuisance parameters characterizing the time-evolution of the progenitor. Without velocity information or a distance gradient along the tidal feature, 4 out of the 6 simulated phase-space dimensions cannot be compared to the measured data directly. Only the on-sky positions can be directly compared,  requiring one to marginalize over the unobserved phase-space dimensions. This marginalization step introduces a range of additional degeneracies in attempting to reconstruct the potential.

\begin{figure*}[tp!]
    \centering
    \includegraphics[scale=.4]{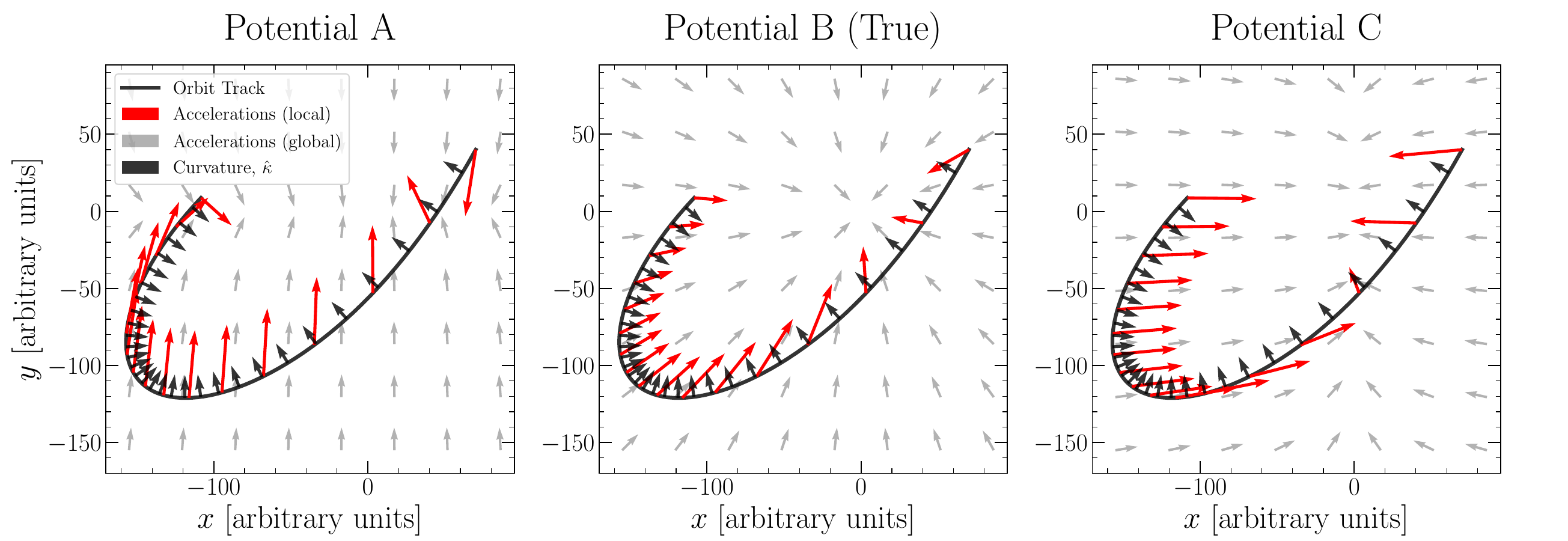}
    \caption{A single orbit and acceleration vectors generated from different potential models. In each panel the same 2D projection of a 3D orbit is shown in black. The curvature vectors, which point orthogonal to the projected orbit, are the black arrows. The gray and red arrows depict acceleration vectors for three different potential models: A, B, and C, all projected onto the $x-y$ plane. Which panel shows the correct acceleration vectors for the given orbit? The correct potential must produce acceleration vectors that fall within $90$~\rm{deg} of the curvature vector. Otherwise, the orbit would not curve in the observed direction. The correct potential is therefore Potential B, since potentials A and C do not satisfy this constraint over the full length of the orbit.}
    \label{fig: potential_guess}
\end{figure*}

 Here, we aim to address what can possibly be learned about the host potential from only the projected track of a tidal stream. To this end, we connect the projected morphology (i.e., curvature) of a tidal feature to the gravitational acceleration field of its host galaxy. The resulting method for potential reconstruction is analytic, adaptable to any static potential model, and does not require trial orbit integrations. We demonstrate the performance of our method on $N-$body simulations of streams, and discuss what morphological properties of streams will be most useful for constraining the potential. 

The paper is organized as follows. In \S\ref{sec: modeling} we introduce our modeling approach. In \S\ref{sec: likelihood} we devise a likelihood. In \S\ref{sec: demonstration_nbody} we apply the method to a series of $N-$body streams generated in potentials with different shape properties and viewing angles. In \S\ref{sec: application_ngc59076} we present preliminary results on the stream surrounding the galaxy NGC 5907. We discuss our results and future prospects in \S\ref{sec: discussion}, and conclude in \S\ref{sec: conclude}.

\section{Modeling Framework}\label{sec: modeling}
We now develop a modeling framework that connects the projected track of a stream to the underlying gravitational potential of its host galaxy. We first consider the intuitive case of orbits, where the local curvature of an orbit is related to the gravitational acceleration field (\S\ref{sec: intuitive_picture}). We then introduce a coherence condition which, if satisfied along the stream track, makes the same orbit-based framework applicable---even if the track of the stream does not trace a single stellar orbit (\S\ref{sec: formal_derivation}). Our main assumptions are summarized below, and we refer the reader to the corresponding sections for further discussion. 
\begin{itemize}
    \item \textbf{Stream Track and Coordinate System (\S\ref{sec: coord_system}):} The observed tidal feature has an elongated morphology in the plane of the sky which can be represented as a twice-differentiable track. 
    
    \item \textbf{Coherence Condition (\S\ref{sec: curvature_accelerations}):} The acceleration direction of a stream ``fluid-element" in projection must not oppose the observed curvature direction of the projected stream-track. 
    Otherwise, the orbits of stars populating the stream would completely decouple from the observed track. We derive an exact inequality that characterizes when this coherence condition is satisfied, and show that it does not require the stream to trace a single stellar orbit. 
    
    \item \textbf{Absence of Curvature (\S\ref{sec: zero_curve}):} If the observed stream has a linear segment with zero projected curvature, we assume that the stream-segment is not accelerating perpendicular to its elongated axis in the plane of the sky. 
\end{itemize}

\subsection{Intuitive Picture: A Single Orbit}\label{sec: intuitive_picture}
In each panel of Fig.~\ref{fig: potential_guess} we plot the same 2D projection of an orbit generated in a 3D potential $\Phi(x,y,z)$. The red and gray vector fields represent accelerations under three logarithmic potential potential models (A, B, C) with different flattening parameters.  The vector field for the correct potential is shown in the middle panel (Potential B). The black arrows indicate unit curvature vectors, which characterize the local concavity of the orbit. Unit curvature vectors are calculated locally, by differentiating the track of the orbit (black curve) in the $x-y$ plane. For illustration purposes we assume a fixed trial distance $z_0$ when plotting all acceleration vectors, though our formal analysis considers a range of possible distances.

Fig.~\ref{fig: potential_guess} illustrates a connection between the curvature of the orbit in projection and the underlying acceleration field. Namely, under the correct potential model, acceleration vectors in the $x-y$ plane evaluated at the location of the orbit always have a component along the local curvature vector. Otherwise, the orbit would not have curved in the observed direction. 
We mathematically derive this to be the case in the following section, though this simple principle forms the core of our analysis. In general, the projection of the acceleration vector along an orbit must always point within $90~\rm{deg}$ of the projected curvature vector. If the orbit has no curvature in projection then we are viewing the orbital plane edge on, and all accelerations are along the line of sight.

If the 3D track of the orbit is known along with its speed, then the acceleration can be calculated directly \citep{Nibauer}. If the speed is not known and the orbit is assumed to be circular, then the correct acceleration vector must point in the same direction as the curvature vector. However, if only a random 2D-projection of the orbit is observed, then the topology of the orbit can be obscured by projection effects. For instance, circles can appear elliptical depending on their orientation with respect to the observer, and ellipses can appear more circular. Without knowing the speed, and without imposing additional priors (e.g., about the potential, circular or radial orbits, projection effects), we can only make the minimal assumption that the projected acceleration field along the orbit must point within  $90$~\rm{deg} of the orbital curvature vector in the plane of the sky. 

This principle can be used to rule out acceleration vectors that could not have possibly generated the observed curvature. For a given potential model, the direction of the acceleration field in a region is sensitive to shape parameters, e.g., how the mass distribution is oriented, flattened, or elongated. We devise a likelihood to perform this inference in \S\ref{sec: likelihood}. Note that the direction of acceleration vectors is not sensitive to the total mass of the system, since uniformly scaling the potential by a constant changes only the magnitude of accelerations. Because of this, orbital curvature alone cannot be used to determine an absolute mass estimate. However, varying the mass ratio between distinct potential components with different shapes (e.g, a disk and halo) does change the direction of accelerations, and therefore orbital curvature. We later show that limits on disk-to-halo mass ratios can be placed, assuming that the disk's potential is aligned with its stellar mass in shape and extent (\S\ref{sec: mass_ratios}). 

There are several caveats to the discussion above. For instance, streams are not orbits \citep{2013MNRAS.433.1813S}. Instead, streams are populated by stars on an ensemble of orbits, with a typically small, albeit non-zero energy gradient along the stream (e.g., \citealt{1998ApJ...495..297J,2001ApJ...557..137J}). Furthermore, the gravitational potential of the Milky Way and external galaxies is not expected to be static. This can affect the orbit of stars in a stream (e.g., \citealt{2019MNRAS.487.2685E, 2021ApJ...923..149S,2023MNRAS.518..774L}). We next consider under what conditions the curvature of a stream can be linked to the acceleration field of its host. The resulting principle is the same as the orbit-based picture we have already discussed, though in \S\ref{sec: formal_derivation} we show that this picture can be extended beyond the case of a single orbit.

\subsection{Formal Derivation: Connecting Stream Curvature to the Gravitational Acceleration Field}\label{sec: formal_derivation}
In the remainder of this section we model streams in projection, and consider what can be learned from the observed shape of a stream without making restrictive assumptions about its speed or shape along the line-of-sight. We show that the curvature-acceleration connection does not require the stream track to trace an orbit (i.e., an isoenergy curve).

\subsubsection{Defining the Stream Track in On-Sky Coordinates}\label{sec: coord_system}
\begin{figure}[tp!]
    \centering
    \includegraphics[scale=.9]{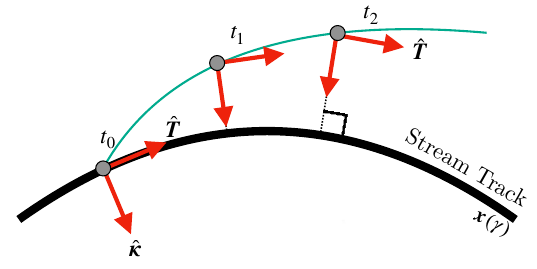}
   \caption{Time-evolution of a stream-element (gray) from the present day $(t_0)$ to a future time $(t_2)$. As the stream element evolves, its basis vectors are set by the minimum projected distance between the element's position and the stream track (black curve, $\boldsymbol{x}(\gamma)$). Tangent ($\hat{\boldsymbol{T}}$) and curvature ($\hat{\boldsymbol{\kappa}}$) unit vectors are set by the morphology of the stream track. }
    \label{fig: Stream_Track_Evo}
\end{figure}
In this section we define the stream track in on-sky coordinates, and decompose the velocity of a stream ``fluid element" into motion along the track and perpendicular to the track in the plane of the sky. We then write the acceleration of the fluid element in this frame, drawing a connection between projected stream-curvature and projected accelerations.

We consider streams that have a well-defined track, e.g., those resembling a series of loops rather than shells. 
We work in a coordinate system centered on the external galaxy where $x-y$ is the plane of the sky and the $z$-axis is along the line-of-sight. The origin is at the center of the host galaxy.

Position along the projected stream-track is described by the vector-valued function $\boldsymbol{x}(\gamma)\in\mathbb{R}^2$, where $\gamma$ is a dimensionless phase-parameter encoding position along the stream. For simplicity, we work in physical coordinates so that  $\boldsymbol{x}(\gamma)$ has units of length. This is not strictly necessary, since our acceleration analysis is scale-free. Instead, $\boldsymbol{x}(\gamma)$ could also represent pixel-based coordinates, or on-sky position angles (e.g., $\phi_1,\phi_2$). 

To characterize the morphology of a stream, we describe the stream track with a series of tangent vectors and orthogonal curvature vectors. The unit tangent vector is defined as
\begin{equation}
    \hat{\boldsymbol{T}}(\gamma) \equiv \frac{d{\boldsymbol{x}}/d\gamma}{\Vert d{\boldsymbol{x}}/d\gamma\Vert},
\end{equation}
and the unit-curvature vector is
\begin{equation}
    \hat{\boldsymbol{\kappa}}(\gamma) \equiv \frac{d\hat{\boldsymbol{T}}/d\gamma}{\Vert d\hat{\boldsymbol{T}}/d\gamma\Vert}.
\end{equation}
These vectors are orthogonal, satisfying $\hat{\boldsymbol{T}}\cdot\hat{\boldsymbol{\kappa}} = 0$. An illustration of the coordinate system is shown in Fig.~\ref{fig: Stream_Track_Evo}. The stream-track is depicted by the solid black curve, and the time-evolution of a stream element is shown in gray. Importantly, $\hat{\boldsymbol{\kappa}}$ is invariant to the circulation direction along the track: it does not matter which tail is ``leading" or ``trailing" in our analysis. 

We now consider the motion of the gray stream element in Fig.~\ref{fig: Stream_Track_Evo}, and under what conditions its acceleration vector is constrained by the curvature vector $\hat{\boldsymbol{\kappa}}$. At the present epoch, $t_0$, the velocity of the element in the plane of the sky is
\begin{equation}\label{eq: velo}
    \boldsymbol{v}_{\rm{planar}} = v_T \hat{\boldsymbol{T}} + v_\kappa \hat{\boldsymbol{\kappa}},
\end{equation}
where $v_T$ and $v_\kappa$ are the tangent and perpendicular velocity components, respectively. For simplicity, we assume that $v_T > 0$.
\footnote{Because we will only constrain perpendicular accelerations, the actual sign of $v_T$ for a given circulation direction is ambiguous and only a matter of bookkeeping.} If the stream traces an orbit, then the stream element will evolve purely along the stream track with $v_\kappa = 0$. More realistically, the stream element could have $v_\kappa \neq 0$. As the stream element evolves in time, the vectors $\hat{\boldsymbol{T}}$ and $\hat{\boldsymbol{\kappa}}$ are chosen by the minimum projected distance between the present-day stream track and the position of the stream element (dashed lines in Fig.~\ref{fig: Stream_Track_Evo}). 

As a stream-element time-evolves, $\boldsymbol{v}_{\rm planar}$ and its components will change according to the planar acceleration 
\begin{equation}\label{eq: a_xy}
\begin{split}
    \boldsymbol{a}_{\rm{planar}} &\equiv  \frac{d\boldsymbol{v}_{\rm{planar}}}{dt} \\ &= \underbrace{\left[\dot{v}_T \hat{\boldsymbol{T}} + v_\kappa \frac{d\hat{\boldsymbol{\kappa}}}{dt} \right]}_{\boldsymbol{a}_T = (\boldsymbol{a}_{\rm{planar}} \cdot \hat{\boldsymbol{T}}) \hat{\boldsymbol{T}}} + \underbrace{\left[\dot{v}_\kappa \hat{\boldsymbol{\kappa}} + v_T \frac{d\hat{\boldsymbol{T}}}{dt}\right]}_{\boldsymbol{a}_\perp =( \boldsymbol{a}_{\rm{planar}} \cdot \hat{\boldsymbol{\kappa}}) \hat{\boldsymbol{\kappa}}},
\end{split}
\end{equation}
where ``over-dot" indicates a time-derivative, and  $\boldsymbol{a}_T$ and $\boldsymbol{a}_\perp$ are the tangential and perpendicular accelerations in the plane of the sky, respectively. By construction, $d\hat{\boldsymbol{T}}/dt$ will point along $\hat{\boldsymbol{\kappa}}$, and $d\hat{\boldsymbol{\kappa}}/dt$ will point along $- \hat{\boldsymbol{T}}$. The latter is sensitive to the circulation direction, so we do not use $\boldsymbol{a}_T$ to constrain the potential. In \S\ref{sec: curvature_accelerations}, we discuss how and when Eq.~\ref{eq: a_xy} can be used to constrain properties of the potential using the morphology of a stream.  

In subsequent sections we constrain $\boldsymbol{a}_{\rm planar}$ through a gravitational potential $\Phi_{\boldsymbol{m}}\left(x,y,z\right)$ with model parameters $\boldsymbol{m}$.
The coordinates $(x,y)$ are on-sky position coordinates, and $z$ is the unknown line-of-sight coordinate. The acceleration is related to the potential through its spatial gradient
\begin{equation}\label{eq: a_planar_phi}
    \boldsymbol{a}\left(x,y,z\right) = -\nabla \Phi_{\boldsymbol{m}}\left(x,y,z\right),
\end{equation}
where $\boldsymbol{a} = \left(a_x,a_y,a_z\right)$. In these coordinates, the planar acceleration vector is
\begin{equation}\label{eq: a_planar_cart}
    \boldsymbol{a}_{\rm planar}\left(x,y,z\right) = a_x\left(x,y,z\right)\hat{\boldsymbol{x}} + a_y\left(x,y,z\right)\hat{\boldsymbol{y}}.
\end{equation}
 
\subsubsection{Connecting Curvature to Accelerations}\label{sec: curvature_accelerations}
We now use Eq.~\ref{eq: a_xy} to express $\boldsymbol{a}_\perp$ in terms of the observable stream curvature, $\hat{\boldsymbol{\kappa}}$. The unit curvature vector is related to the time derivative of the unit tangent vector through
\begin{equation}\label{eq: kappa_def}
    \hat{\boldsymbol{\kappa}}\left(\gamma(t)\right) = \frac{d\boldsymbol{\hat{T}}/dt}{\Vert d\boldsymbol{\hat{T}}/dt \Vert } = \frac{d\hat{\boldsymbol{T}}/d\gamma}{\Vert d\hat{\boldsymbol{T}}/d\gamma \Vert},
\end{equation}
where $\gamma$ is the dimensionless phase-parameter, parameterizing the track of the stream. Eq.~\ref{eq: kappa_def} allows us to write an expression for $\boldsymbol{a}_\perp$ (Eq.~\ref{eq: a_xy}) in terms of the observed unit-curvature vector along the stream track, namely,
\begin{equation}\label{eq: a_perp_redef}
    \boldsymbol{a}_\perp = \left( \dot{v}_\kappa + v_T \Big\Vert \frac{d\hat{\boldsymbol{T}}}{dt} \Big\Vert \right)\hat{\boldsymbol{\kappa}}.
\end{equation}
By construction, $\boldsymbol{a}_\perp$ points along $\pm \hat{\boldsymbol{\kappa}}$: the sign is determined by the relative size between the two terms in parentheses. We use Eq.~\ref{eq: a_perp_redef} to write a \emph{coherence condition}, which specifies precisely when $\boldsymbol{a}_\perp$ point along $+\hat{\boldsymbol{\kappa}}$. Namely, 
\begin{equation}\label{eq: require}
\Big\Vert \frac{d\hat{\boldsymbol{T}}}{dt}\Big\Vert > -\frac{\dot{v}_\kappa}{v_T}.
\end{equation}
When Eq.~\ref{eq: require} is satisfied, $\boldsymbol{a}_\perp$ points along $+\hat{\boldsymbol{\kappa}}$ and the observed stream-curvature provides a constraint on the acceleration direction. 

We now discuss whether we expect Eq.~\ref{eq: require} to hold for streams. In static potentials, stream stars tend to sort out by their energies (see, e.g., \citealt{2001ApJ...557..137J}), so that proximate regions of the stream are populated by stars on similar orbits without large oscillations away from the stream track. This behavior is seen in action-angle coordinates, where stars tend to sort out in angle by their frequency difference relative to the progenitor \citep{2013MNRAS.433.1813S}. Then, at a given phase, the stream is populated by an ensemble of stars with locally similar frequencies. In this case, we do not expect stars to suffer large oscillations away from the stream track. Then, Eq.~\ref{eq: require} would be satisfied. This property of streams has been utilized to measure the solar velocity \citep{2020arXiv201205271M}, and to discover new streams in the Milky Way \citep{2021ApJ...914..123I}. 

Eq.~\ref{eq: require} is not a particularly strict requirement: if $\dot{v}_\kappa << 0$, the stream track would be totally decoupled from its trajectory, and the stream would begin the process of phase-mixing again from its new configuration. The mere existence of a stream implies a level of coherence, though time-dependent potentials can lead to perpendicular motion to the stream track \citep{2019MNRAS.487.2685E,Vasiliev_2020}. This does not necessarily violate Eq.~\ref{eq: require}, provided that $\dot{v}_\kappa$ is bounded below by $-v_T \Vert d\hat{\boldsymbol{T}}/dt\Vert$.

As long as Eq.~\ref{eq: require} is satisfied, the stream-element's orbital trajectory has a perpendicular acceleration component along $\hat{\boldsymbol{\kappa}}$. This means
\begin{equation}\label{eq: perp_acc_unit_kappa}
    \frac{\boldsymbol{a}_\perp}{\Vert \boldsymbol{a}_\perp \Vert} = \hat{\boldsymbol{\kappa}},
\end{equation}
where $\hat{\boldsymbol{\kappa}}$ can be estimated from data directly (Eq.~\ref{eq: kappa_def}). 

If the stream track were to trace an orbit, then $v_\kappa = \dot{v}_\kappa = 0$. However, the connection between perpendicular accelerations and stream-curvature (Eq.~\ref{eq: perp_acc_unit_kappa}) holds even for $\dot{v}_\kappa \neq 0$. Thus, the track of the stream does not need to actually trace an orbit in order to connect stream-curvature to perpendicular accelerations. 

\begin{figure}[tp!]
    \centering
    \includegraphics[scale=0.95]{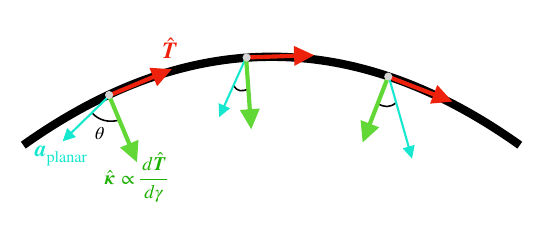}
    \caption{Illustration of a stream track (black curve), its tangent vectors ($\hat{\boldsymbol{T}}$), and curvature vectors ($\hat{\boldsymbol{\kappa}}$). The planar accelerations along the track are the cyan vectors, ${\boldsymbol{a}}_{\rm{planar}}$. The angle between $\hat{ {\boldsymbol{\kappa}} }$ and ${\boldsymbol{a}}_{\rm{planar}}$ is denoted with $\theta$, which depends on the underlying gravitational potential, $\Phi$.}
    \label{fig: Acc_Normal_Angle}
\end{figure}

We now cast Eq.~\ref{eq: perp_acc_unit_kappa} as a geometric constraint, which will be useful in interpreting our results. From Eq.~\ref{eq: a_xy}, $\boldsymbol{a}_\perp = (\boldsymbol{a}_{\rm{planar}} \cdot \hat{\boldsymbol{\kappa}}) \hat{\boldsymbol{\kappa}}$. The unit vector of $\boldsymbol{a}_\perp$ is then
\begin{equation}\label{eq: a_perp_over_mag}
\begin{split}
    \frac{\boldsymbol{a}_\perp}{\Vert \boldsymbol{a}_\perp \Vert} &= \frac{\boldsymbol{a}_{\rm{planar}} \cdot \hat{\boldsymbol{\kappa}}}{\Vert {\boldsymbol{a}_{\rm{planar}} \cdot \hat{\boldsymbol{\kappa}}} \Vert } \hat{\boldsymbol{\kappa}} \\
    &= \frac{\cos{\left(\theta( \Phi) \right)}}{\vert \cos{\left(\theta( \Phi) \right)} \vert} \hat{\boldsymbol{\kappa}} \\
    &= \sgn{\left[ \cos{\left(\theta( \Phi) \right)} \right]}\hat{\boldsymbol{\kappa}},
\end{split}
\end{equation}
where $\sgn(x)$ is the sign function, with $\sgn{(x>0)} = +1$, $\sgn{(x=0)} = 0$, and $\sgn{(x<0)} = -1$. The symbol $\theta(\Phi) \in [-\pi, \pi]$ represents the angle between the planar acceleration, $\boldsymbol{a}_{\rm{planar}}$, and the curvature vector to the stream track, $\hat{\boldsymbol{\kappa}}$. Planar accelerations depend on the underlying gravitational potential, $\Phi$. The angle $\theta\left(\Phi\right)$ is illustrated in Fig.~\ref{fig: Acc_Normal_Angle}, where the cyan arrows depict planar acceleration vectors along the stream track, and the green arrows are unit curvature vectors along the track $(\hat{\boldsymbol{\kappa}})$. When the coherence condition expressed in Eq.~\ref{eq: require} is satisfied, $\boldsymbol{a}_\perp$ points along $\hat{\boldsymbol{\kappa}}$ and  $\sgn{\left[ \cos{\left(\theta( \Phi) \right)} \right]} = 1$. This implies 
\begin{equation}\label{eq: delta_theta_constraint}
    \vert \theta\left(\Phi\right) \vert < \pi/2,
\end{equation}
or that the angle between $\boldsymbol{a}_{\rm planar}$ and $\hat{\boldsymbol{\kappa}}$ will fall within $\pi/2$ radians. 

Intuitively, Eq.~\ref{eq: delta_theta_constraint} requires that a change in the unit tangent vector along the projected stream track must have been induced by an acceleration perpendicular to the track. The direction of this planar acceleration must fall within $90~\rm{deg}$ of the curvature vector to produce the stream's morphology (Eq.~\ref{eq: delta_theta_constraint}). In the subsequent sections, we explore what limits this basic constraint, combined across the length of a stream, can place on the underlying gravitational potential. 

The analysis presented here uses only the curvature direction and not curvature strength to constrain planar accelerations. While the latter does depend on the acceleration field, it also depends on the velocity of the stream (e.g., $\Vert d\hat{\boldsymbol{T}}/dt \Vert$). In this work we aim to address what can be inferred from only the track of the stream without restrictive velocity assumptions. However, priors can be placed on the velocity of each stream segment to provide even tighter constraints. We leave this to future work.

\subsubsection{Special Case: Zero Curvature}\label{sec: zero_curve}

We have previously assumed that the projected stream track is non-linear. However, when $\Vert d\hat{\boldsymbol{T}}/d\gamma \Vert \approx 0$, the stream-track is locally linear in the $x-y$ plane. This could correspond to an inflection point where the stream changes concavity, or a radial orbit. We take a lack of curvature to indicate that the perpendicular acceleration component, $\boldsymbol{a}_\perp$, is equal to zero. We justify this below. 

 If we assume that the coherence condition in Eq.~\ref{eq: require} is satisfied, then at an inflection point where the stream changes concavity it must be the case that $\boldsymbol{a}_\perp = 0$. This is required in order to satisfy Eq.~\ref{eq: require} on either side of the inflection point, provided that $\boldsymbol{a}_\perp$ is continuous along the stream. The result is that at an inflection point, $\boldsymbol{a}_\perp = 0$ and $\boldsymbol{a}_{\rm planar}$ points entirely along $\pm \hat{\boldsymbol{T}}$ for $\mathbf{a}_{\rm planar} \neq 0$.  


Alternatively, a linear stream segment in the plane of the sky could haven been observed at an epoch when the stream was undergoing a transition from having curvature in one direction, to zero curvature, and finally to curvature in the opposite direction (that is, from ``u-shaped", to linear, to ``n-shaped"). We assume that such a scenario is unlikely to occur, and treat extended segments of a stream with zero curvature as having $\boldsymbol{a}_\perp = 0$.

In practice, due to statistical uncertainty in our estimation of the stream track, the derivative $\Vert d\hat{\boldsymbol{T}}/d\gamma \Vert$ will not evaluate to exactly 0 even for streams with negligible curvature. Instead, the 
angular change in the tangent vector $\hat{\boldsymbol{T}}$ over an increment in angular arc-length, $d\ell$, is the dimensionless number: 
\begin{equation}
\Big\vert \frac{d\phi}{d\ell}\Big\vert = \Big\Vert \frac{d\hat{\boldsymbol{T}}}{d\gamma} \Big\Vert \left(\frac{d\ell}{d\gamma} \right)^{-1},
\end{equation}
where $\phi$ is (e.g.) the angle between the tangent vector, $\hat{\boldsymbol{T}}$, and a fixed reference axis in the plane of the sky. In Appendices~\ref{sec: fitting_routine}-\ref{app: Hyper_Params} we discuss the stream track fitting routine and the threshold $\epsilon$, which describes when we treat a stream segment as linear ($\vert d\phi/d\ell \vert < \epsilon$).

\section{Statistical Model}\label{sec: likelihood}
\subsection{Overview}\label{sec: overview}
In this section we develop a likelihood to apply the formalism developed in \S\ref{sec: modeling} to observational data. For our data inputs, we measure curvature unit vectors from a projected stream track that has been fit to the data (i.e., from imaging of the stream). We assume that $N$ such vectors have been estimated from the data, and make up the set $\{\hat{\boldsymbol{\kappa}}_i\}_{i=1}^N$. The model component of the likelihood is the acceleration field evaluated along the stream, generated from a gravitational potential $\Phi_{\boldsymbol{m}}(x,y,z)$ with model parameters $\boldsymbol{m}$ (Eq.~\ref{eq: a_planar_phi}). The coordinates $(x,y)$ are on-sky position coordinates, and $z$ is the unknown line-of-sight coordinate. In our approach the acceleration field is allowed to vary along the line of sight, even in the absence of distance information along the stream. We handle the unknown line-of-sight coordinate by maximizing the likelihood over all distance tracks (discussed in \S\ref{sec: distance_gradients}).

\subsection{Dealing with Unobserved Line-of-Sight Distances}\label{sec: distance_gradients}
In general, planar accelerations depend on both the on-sky position coordinates $(x,y)$ and the line-of-sight coordinate, $z$. As a result, the angle between the stream curvature vectors and the planar component of the acceleration vector depends on the 3D location of the stream segment. We now discuss how our curvature-acceleration analysis treats the unobserved line-of-sight coordinate, by maximizing the likelihood over all possible continuous distance tracks in $z$. Then, the best possible potential model satisfies $\vert \theta_i \vert < \pi/2$  for all $i$ while allowing for the possibility of a connected stream.  This constraint is expressed mathematically in \S\ref{sec: likelihood_specific}, though we provide a more qualitative description here. 

In practice, each evaluation point along the stream track has some range of distances that yield $\vert \theta_i\vert<\pi/2$. For the $i^{\rm th}$ evaluation point along the stream track, let $D_i$ represent the set of line-of-sight distances from the host galaxy in the interval $[z_{\rm min}, z_{\rm max}]$ for which $\vert \theta_i \vert < \pi/2$. Throughout this work we set $z_{\rm max} = -z_{\rm min} \approx 100~\rm{kpc}$. For some potential models we might find that the allowed distance ranges for subsequent stream segments (i.e., $D_{i-1}$ and $D_{i}$) do not overlap. However, we expect distance to vary smoothly along the stream. This means that subsequent stream segments should have overlapping distance ranges within some reasonable threshold, $\Delta D$. We impose this as a prior, by requiring that $D_i$ and $D_{i-1}$ overlap within $\Delta D$. We choose $\Delta D = 30~\rm{kpc}$, since this is a conservative upper-limit on the distance between subsequent evaluation points along a stream.

The distance sampling scheme has a few ``edge cases," addressed in Appendix~\ref{app: likelihood_edge}. From the above discussion, it is evident that our analysis could, in principle, constrain the 3D shape of a stream track from only its projected morphology. Typically, we find that 3D tracks are poorly constrained, with the exception of a few special cases. We postpone exploring constraints on 3D stream tracks (i.e., measuring distance gradients) to a future work.

If we view the total mass distribution along a principal axis (\S\ref{sec: viewed_along_symm}), then planar accelerations vary only in magnitude along the line of sight. In this case, all distance tracks are valid and requiring the stream to have a continuous distribution in $z$ does not affect the likelihood. 

\subsection{Likelihood}\label{sec: likelihood_specific}
We now develop a likelihood that connects the model introduced in \S\ref{sec: modeling} to the data. Our strategy is to evaluate the likelihood of the observed stream curvature, given the planar acceleration field of a potential model with parameters $\boldsymbol{m}$. We construct the likelihood to prefer curvature-acceleration angles within $ 90~\rm{deg}$ (i.e., planar acceleration vectors pointing inside the stream's curve), and a vanishing perpendicular acceleration at linear segments of a stream.

Because we constrain the acceleration field along the stream track, we must also consider where in 3D space the potential model is evaluated. For the $i^{\rm th}$ evaluation point along the stream track, the on-sky coordinates are $(x(\gamma_i),  y(\gamma_i))$. Let $z(\gamma)$ represent the unknown distance track of the stream. The full 3D position vector is $\boldsymbol{X}(\gamma)$. For a trial distance track, we consider the three cases:
\begin{itemize}
    \item {\it Case 1} ($C_1$): $\vert \theta_i \vert < \pi/2$ at $\boldsymbol{X}(\gamma_i)$.
    \item {\it Case 2} ($C_2$): $\vert \theta_i \vert > \pi/2$ at $\boldsymbol{X}(\gamma_i)$.
    \item {\it Case 3} ($C_3$): The stream segment has negligible projected curvature and $\theta_i$ is undefined. 
\end{itemize}
Each evaluation point at $\boldsymbol{X}(\gamma_i)$ corresponds to one of the three cases. This allows us to write the probability for the $i^{\rm th}$ curvature vector, conditioned on the line-of-sight coordinate $z(\gamma_i)$, model parameters $\boldsymbol{m}$, and the specific case $C_j$. For the first two cases, we write this conditional probability as
\begin{equation}\label{eq: C12_likelihood}
    P\left(\hat{\boldsymbol{\kappa}}_i \Big|   \boldsymbol{m}, z(\gamma_i), C_{\{1,2\}} \right) \propto \Bigg\{
    \begin{array}{lr}
        1 \ \rm{if} \ C_{\{1,2\}} \ \rm{is \ satisfied}\\
        0 \ \rm{otherwise}
    \end{array}
\end{equation}
where $C_{\{1,2\}}$ can be replaced with either $C_1$ or $C_2$. Note that ``otherwise" in Eq.~\ref{eq: C12_likelihood} is applicable if $C_{\{1,2\}}$ is not satisfied, or if $\hat{\boldsymbol{\kappa}}_i$ is undefined in the absence of curvature. The proportionality sign is sufficient, since the normalizing constant is an overall scaling factor in the total likelihood. We treat Eq.~\ref{eq: C12_likelihood} as a binary outcome, since the curvature either is or is not compatible with the given condition ($C_1$ or $C_2$). 

For condition $C_3$, the stream appears locally linear in projection and $\hat{\boldsymbol{\kappa}}$ is undefined. In this case, rather than enforcing that $\boldsymbol{a}_{\rm planar}$ points entirely along $\pm \hat{\boldsymbol{T}}$ (\S\ref{sec: zero_curve}), we allow for deviations from this behavior by placing a Gaussian prior on the angle between $\boldsymbol{a}_{\rm planar}$ and the stream track. We adopt this approach since any given stream will have a non-negligible thickness, making it difficult to determine precisely what straight line path should be traced to construct the stream track. Placing a Gaussian prior on the angle between the planar acceleration vector and stream track in this case is more conservative, since we will not rule out potential models that produce the approximately correct behavior (i.e., planar accelerations nearly aligned with $\pm \hat{\boldsymbol{T}}$). The angle between the planar acceleration vector and the stream track is given by
\begin{equation}\label{eq: theta_t}
    \theta_{T,i} \equiv \pi/2 - \arccos{\left( \hat{\boldsymbol{a}}_{\rm{planar},i} \cdot \hat{\boldsymbol{N}}_i \right)},
\end{equation}
where $\hat{\boldsymbol{N}} \equiv \left(-T_y, T_x\right)$ is orthogonal to $\hat{\boldsymbol{T}} = \left(T_x, T_y\right)$ in the plane of the sky. This definition ensures $\theta_{T,i} \in [-\pi/2, \pi/2]$. Note that $\hat{\boldsymbol{a}}_{{\rm planar},i}$ depends on $z$, so $\theta_{T,i}$ also depends on the line-of-sight distance along the stream track. In this case, we write the conditional probability for a stream element with negligible curvature as 
\begin{multline}\label{eq: NoCurve_likelihood}
    P\left( \hat{\boldsymbol{\kappa}}_i |\boldsymbol{m}, z(\gamma_i),  C_3 \right) \propto \\ \Bigg\{
    \begin{array}{lr}
         \mathcal{N}\left( \theta_T |\mu = 0, \sigma = \sigma_{\theta_T} \right) \ \rm{if \ } C_3 \ \rm{is \ satisfied} \\
        0 \ \rm{otherwise}
    \end{array}
\end{multline}
where $\mathcal{N}$ is the Gaussian distribution with mean $\mu$ and standard deviation $\sigma$. Eq.~\ref{eq: NoCurve_likelihood} up-weights acceleration vectors that point tangent to the stream track (small $\theta_T$), and down-weights acceleration vectors with large $\theta_T$. In practice, because $\theta_T$ is set to a fixed interval $\mathcal{N}$ should really be a truncated Gaussian. However, because we never sample $\theta_T$ outside of this range, the only difference between $\mathcal{N}$ and the properly truncated Gaussian is a normalization, which is why we have written a proportionality symbol instead of equality in Eq.~\ref{eq: NoCurve_likelihood}. In this work we choose $\sigma_{\theta_T} = 10~\rm{deg}$, representing our prior uncertainty on the direction of the stream track. We discuss this choice of the hyperparameter further in Appendix~\ref{app: Hyper_Params}.    

By marginalizing over each case $(C_1,C_2,C_3)$, the likelihood for a curvature vector, $\hat{\boldsymbol{\kappa}}_i$, conditioned on the gravitational potential model parameters, $\boldsymbol{m}$, and line-of-sight track, $z(\gamma)$, is
\begin{multline}\label{eq: L_i}
    P\left(\hat{\boldsymbol{\kappa}}_i |\boldsymbol{m}, z(\gamma) \right) \\= \sum_{j=1}^3 P\left(\hat{\boldsymbol{\kappa}}_i | \boldsymbol{m}, z(\gamma_i), C_j\right)P\left(C_j | \boldsymbol{m}, z(\gamma) \right),
\end{multline}
where $P\left(C_j | \boldsymbol{m}, z(\gamma) \right)$ is the probability (weight) of each respective case with $\sum_j P\left(C_j | \boldsymbol{m}, z(\gamma) \right) =  1$. We assume that the probability of each respective case is constant along the stream track. This allows us to treat $P\left(C_j | \boldsymbol{m}, z(\gamma) \right) \ \mathrm{for} \ j = 1,2,3$ as three parameters, independent of position along the stream. For the gravitational potential parameters $\boldsymbol{m}$ and distance track $z(\gamma)$, the total likelihood taken over the length of the stream (i.e., as a product of Eq.~\ref{eq: L_i} over all $i$) will be maximized when $P\left(C_j|\boldsymbol{m}, z(\gamma)\right)$ is the fraction of times that each case, $C_j$, occurs along the stream track. For instance, if $C_k$ is satisfied over the stream's length then $P(C_k | \boldsymbol{m}, z(\gamma)) = 1$ and $P(C_{j \neq k}| \boldsymbol{m}, z(\gamma)) = 0$ will produce the most positive total likelihood. If case $C_k$ is satisfied for only a small subset of the evaluation points, then $P(C_k |\boldsymbol{m}, z(\gamma))$ will be the proportion of evaluation points for which $C_k$ is satisfied. Otherwise,  giving higher weight to this case would then suppress the other two cases (since the sum of the weights is $1$), despite the fact that they occur more frequently. The weights that maximize the likelihood are 
\begin{equation}\label{eq: fractions}
    \begin{split}
        f_{1} &\equiv P\left(C_1 | \boldsymbol{m}, z(\gamma) \right) =  \frac{1}{N} \sum\limits_{i=1}^{N_{\kappa}} \frac{1}{2}\Big\vert 1 + \sgn{\left(\hat{\boldsymbol{a}}_{\rm{planar},i} \cdot \hat{\boldsymbol{\kappa}}_i \right)} \Big\vert \\
        f_{2} &\equiv P\left(C_2 |  \boldsymbol{m}, z(\gamma) \right) = \frac{N_\kappa}{N} - f_{1} \\
        f_3 &\equiv P\left(C_3 | \boldsymbol{m}, z(\gamma)\right) = 1 - \left[f_{1} +  f_{2}   \right],
    \end{split}
\end{equation}
where the summation is over the $N_{\kappa}$ curvature vectors with non-zero curvature, and $\boldsymbol{a}_{\rm{planar}}$ depends on parameters $\boldsymbol{m}$ and the distance track $z(\gamma)$. From Eq.~\ref{eq: fractions}, $f_{1}$ is the fraction of evaluation points with compatible curvature vectors and planar accelerations, $f_{2}$ is the fraction of evaluation points with incompatible curvature vectors and planar accelerations, and $f_3$ is the fraction of evaluation points with undefined curvature vectors (this is fixed for each stream track). The term in the summation of Eq.~\ref{eq: fractions} is $1$ if the curvature vector is compatible with the proposed planar acceleration, and $0$ otherwise. 

With each term in Eq.~\ref{eq: L_i} defined, we can now write an objective function---maximized over all possible distance tracks  satisfying the constraints outlined in \S\ref{sec: distance_gradients}---as
\begin{equation}\label{eq: likelihood}
    \mathcal{L}\left(\{\hat{\boldsymbol{\kappa}}_i \} | \boldsymbol{m} \right) \equiv \underset{ z(\gamma) \in \rm{\ Paths} }{ \texttt{max}}\left[ 
    \prod_{i=1}^N P\left(\hat{\boldsymbol{\kappa}}_i |\boldsymbol{m}, z(\gamma) \right)\right].
\end{equation}
Eq.~\ref{eq: likelihood} is not a standard likelihood function due to the maximization. In the frequentist literature, a likelihood of this form is referred to as a ``profile likelihood" \citep{Sprott2000StatisticalII}, since it is maximized over nuisance parameters (the distance track, in this case). 

We remind the reader that we do not explicitly sample over all possible paths in the maximization step of Eq.~\ref{eq: likelihood}. As discussed in \S\ref{sec: distance_gradients}, we compute the likelihood recursively along the stream track, so the range of distances that the $i^{\rm th}$ evaluation point could occupy is near the range of distances compatible with the adjacent evaluation point indexed $i - 1$. The distance sampling procedure is discussed further in Appendix~\ref{app: likelihood_edge}.

Eq.~\ref{eq: likelihood} is degenerate with the $f$ parameters. To break the degeneracies, we require
\begin{equation}
    f_{1} \large> f_{2}.
\end{equation} 
This ensures that maximizing the likelihood (Eq.~\ref{eq: likelihood}) entails finding the acceleration field that is most consistent with the measured curvature vector: i.e., the acceleration field for which Eq.~\ref{eq: delta_theta_constraint} is most frequently satisfied across the stream. 

\section{Tests on Simulated Stellar Streams}\label{sec: demonstration_nbody}
In this section we demonstrate the method on $N-$body streams generated in a ground truth background potential. We first discuss the $N-$body simulations in \S\ref{sec: gen_Nbody}, then test our method using streams generated in an axisymmetric halo-only potential (\S\ref{sec: n_body_results}). In \S\ref{sec: disk_halo_misalignment} we apply the method to streams generated in a more complicated disk + halo potential, and consider measuring the mass ratio of the two components in \S\ref{sec: mass_ratios}. In \S\ref{sec: viewed_along_symm} we assume that the galaxy and its halo are viewed along a symmetry axis, and relax this assumption in \S\ref{sec: 3D_shape}.

\subsection{Simulations}\label{sec: gen_Nbody}
In this section we discuss the details of our simulations, including initial conditions (\S\ref{sec: init_conds}), the potential models considered (\S\ref{sec: potentials}), viewing angles (\S\ref{sec: observing_sims}), and fitting a stream track to simulation snapshots (\S\ref{sec: fitting_stream_track}). 

\subsubsection{Initial Conditions}\label{sec: init_conds}
We simulate stellar streams by following the tidal disruption of satellite dwarf galaxies with self-gravity. The satellites are initialized with a King distribution function \citep{1966AJ.....71...64K} with $W_0 = 3$ and $r_s \approx 1~\rm{kpc}$ using the software package \texttt{AGAMA} \citep{2019MNRAS.482.1525V}. We sample the distribution function to generate $5\times 10^6$ particles which are then evolved forward using the $N-$body code \texttt{PyFalcon}, a python interface for \texttt{gyrfalcON} implemented in the \texttt{NEMO} framework \citep{2002JCoPh.179...27D,1995ASPC...77..398T}. The total mass of each cluster is $10^6~M_{\rm{Sun}}$. For each simulation, the progenitor's location is initialized by randomly sampling a point within a galactocentric spherical radius $r \in [20,100]~\rm{kpc}$. The initial velocity is randomly sampled in 0.3--1$V_c$, where $V_c$ is the local circular velocity. This random sampling procedure gives rise to a diverse range of stream morphologies, providing a useful test-set for analyzing distinct tidal features generated in a known ground truth potential.

\subsubsection{Potentials}\label{sec: potentials}
We generate streams using $N-$body simulations with a static external potential. For a stream in the extended stellar halo of its host galaxy, we expect our constraints to be dominated by the dark matter component of the potential. Therefore, our simulations always include a halo potential, though in \S\ref{sec: disk_halo_misalignment}-\ref{sec: mass_ratios} we also incorporate a Miyamoto-Nagai disk \citep{1975PASJ...27..533M}. We view the disk edge-on, so that its midplane is in the $x-z$ plane, and its perpendicular axis is along the $y$ direction. The disk's potential in our reference frame is
\begin{equation}\label{eq: disk}
    \Phi_{\rm Disk}\left(\boldsymbol{x}\right) = -\frac{GM_{\rm Disk}}{\sqrt{x^2+z^2 + \left(\sqrt{y^2 + b^2} + a \right)^2}}.
\end{equation}
The disk has scale-mass $M_{\rm Disk} = 1.2 \times 10^{10}~M_{\rm Sun}$, scale-length $a = 3~\rm{kpc}$, and scale-height $b = 0.5~\rm{kpc}$. These parameter choices are similar to constraints derived from the Milky Way, though the scale mass is somewhat lower than recent estimates, typically based on Jeans modeling or rotation curve estimates (e.g., \citealt{2009PASJ...61..227S,2014ApJ...794...59K,2017OAst...26...72B,2020A&A...642A..95C}). We have explored changing the disk parameters and find that our conclusions are robust to reasonable variations. 

For the halo we use the logarithmic potential, which has a flat rotation curve, similar to the observations. The functional form is
\begin{equation}\label{eq: log_pot}
    \Phi_{\rm Halo} \left(\boldsymbol{x}^\prime\right) = \frac{1}{2}v_c^2 \ln{\left(r_c^2 + \left(\frac{{x^\prime}}{q_1}\right)^2 +  \left( \frac{y^\prime}{q_2}\right)^2 + {z^\prime}^2  \right)}.
\end{equation}
We have introduced the primed coordinates $\boldsymbol{x}^\prime$, which are aligned with the principal axes of the triaxial halo potential. We relate the prime and unprimed frames in \S\ref{sec: observing_sims}. When expressed in a common coordinate frame, the total potential consisting of a disk and halo is simply $\Phi = \Phi_{\rm Disk} + \Phi_{\rm Halo}$. Note that for an axisymmetric logarithmic potential, the relationship between the potential flattening ($q$) and density flattening ($q_{\rm density}$) is $3(1-q) \approx 1-q_{\rm density}$ for moderate flattening (e.g., \citealt{2008gady.book.....B} pg. 77). 

The parameter $v_c$ in Eq.~\ref{eq: log_pot} is fixed such that the circular velocity at $8~\rm{kpc}$ is roughly $250~\rm{km/s}$, and $r_c = 16~\rm{kpc}$. These parameters are similar to Milky Way potential constraints derived from the GD-1 stream in \cite{2019MNRAS.486.2995M}.  For most simulations we set $q_1 = 1$, in which case we refer to $q_2$ as $q$. We carry out multiple simulations with the flattening parameter, $q_2$, varying in the range $[0.75,1.5]$. In \S\ref{sec: triaxial_potential} we attempt to constrain the triaxial potential in Eq.~\ref{eq: log_pot} with $q_1 \neq 1$.

\subsubsection{Observing our Simulations}\label{sec: observing_sims}
In our experiments, the primed frame is aligned with the principal axes of the assumed halo potential, whereas the unprimed frame describes our reference frame. Both frames have the same origin, at the center of the host galaxy. In our reference frame $x-y$ is the plane of the sky, and $z$ is the line-of-sight coordinate.

\begin{figure}[tp!]
    \centering
    \includegraphics[scale=.35]{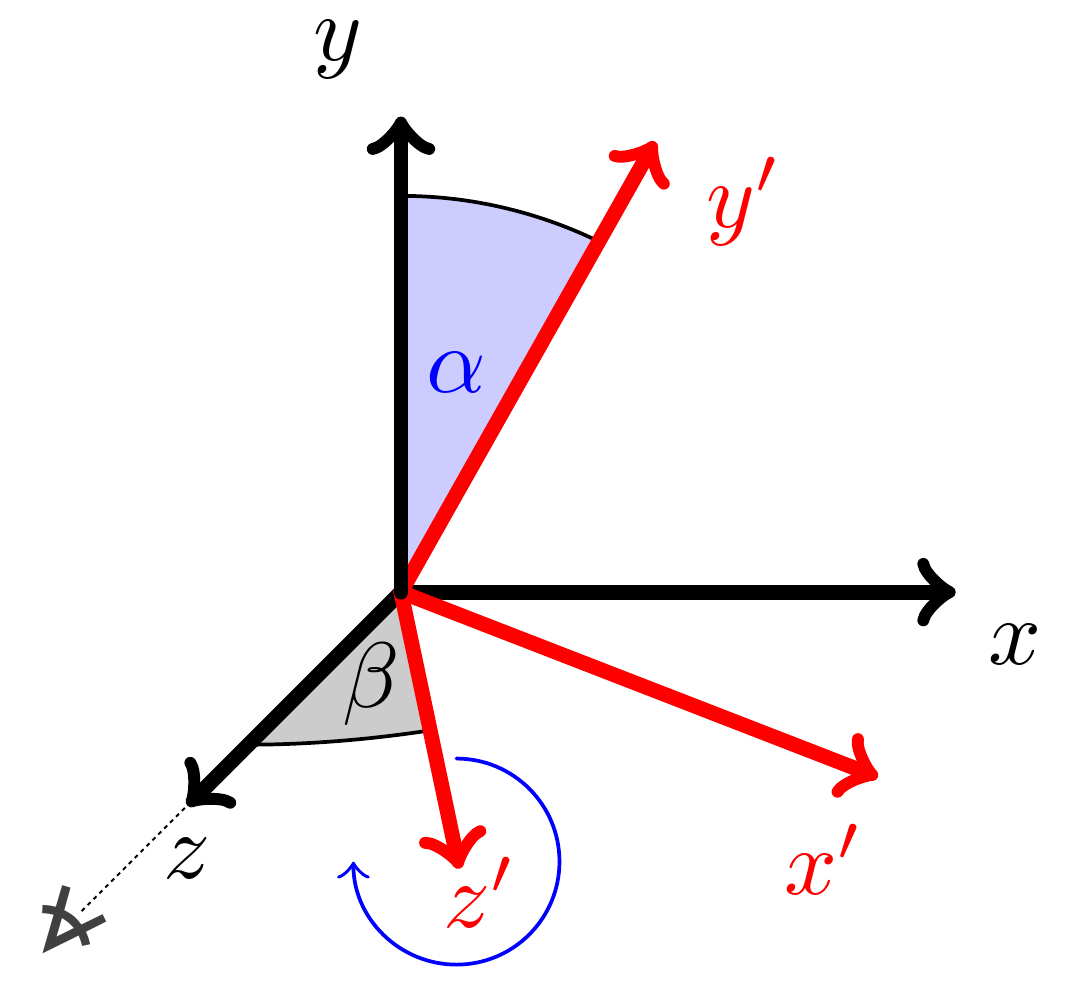}
    \caption{Illustration of the coordinate frames. The $x-y$ plane is the plane of the sky, and the $z$-axis is along the line-of-sight. The primed frame (red) is rotated with respect to the unprimed frame (black), and is aligned with the principal axes of the dark matter halo potential. The angle $\beta$ describes a rotation about the $y$-axis, while $\alpha$ describes a rotation about the $z^\prime$ axis.}
    \label{fig: coord_frame}
\end{figure}

We relate the primed and unprimed frames through the linear transformation
\begin{equation}\label{eq: lin_transf}
    \boldsymbol{x}^\prime = \boldsymbol{R}\boldsymbol{x},
\end{equation}
where $\boldsymbol{R}$ is the rotation matrix
\begin{equation}\label{eq: rot_mat}
    \boldsymbol{R} \equiv \begin{pmatrix}
    \cos{\alpha}\cos{\beta} & \sin{\alpha} &\cos{\alpha}\sin{\beta} \\
    
    -\sin{\alpha}\cos{\beta} &  \cos{\alpha} & -\sin{\alpha}\sin{\beta} \\
    -\sin{\beta}  & 0 & \cos{\beta}
    \end{pmatrix}.
\end{equation}
The matrix $\boldsymbol{R}$ characterizes a counterclockwise rotation about the $y$-axis by an angle $\beta$, followed by a clockwise rotation about the (new) $z^\prime$ axis by an angle $\alpha$ (similar to Euler angles). The relationship between the prime and unprimed frames is illustrated in Fig.~\ref{fig: coord_frame}. We test the impact of projection effects on our inference by observing the same halo model at different viewing angles, defined by $\alpha$ and $\beta$ (\S\ref{sec: 3D_shape}). 

In \S\ref{sec: disk_halo_misalignment} we attempt to constrain a disk-halo misalignment angle, which is the angle between the flattening axis of an axisymmetric halo potential (the $y^\prime$ axis in Eq.~\ref{eq: log_pot}) and an axis orthogonal to the disk's midplane. For the simulations in \S\ref{sec: disk_halo_misalignment}, the disk-halo angle is simply $\alpha$ in Fig.~\ref{fig: coord_frame}, with $\beta = 0$. We use the same definition in \S\ref{sec: two_comp}, where we apply our method to the stream surrounding the edge-on galaxy NGC 5907.

In Table~\ref{table: sim_properties} we provide a summary of the simulations utilized in each section. Multiple simulations with different flattening parameters are indicated by the curly brackets $\{\cdot\}$. For simulations with a disk, the disk column is labeled Y. Otherwise, the column is blank. 
\begin{table}[tp]
\begin{center}
Potential Models
\begin{tabular}{p{1.5cm}p{.8cm}p{1.9cm}p{.8cm}p{.5cm}p{0.5cm}}
\hline\hline
Section & $q_1$ & $q_2$ & Disk & $\alpha$ & $\beta$ \\
\hline
\ref{sec: n_body_results} & 1 & \{0.8, 1, 1.5\} & N & 0$^\circ$ & 0$^\circ$ \\
\hline
\ref{sec: disk_halo_misalignment}-\ref{sec: mass_ratios} & 1 & 0.75 & Y & 30$^\circ$ & 0$^\circ$ \\
\hline
\ref{sec: mass_ratios} & 1 & 1.25 & Y & 0$^\circ$ & 0$^\circ$ \\
\hline
\ref{sec: axisymm_general} & 1 & 1.3 & N & 150$^\circ$ & 85$^\circ$ \\
\hline
\ref{sec: triaxial_potential} & 0.85 & 1.3 & N & 0$^\circ$ & 40$^\circ$ \\
\hline
\ref{sec: what_makes_informative} & 1 & $\{0.8, 1, 1.3\}$ & N & 0$^\circ$ & 0$^\circ$\\
\hline\hline 
\end{tabular}\end{center}
\caption{Parameters of the simulations analyzed in each section. The parameters $q_1$ and $q_2$ characterize the flattening of the halo potential, and the disk column indicates whether or not a disk is included (Y is yes, N indicates no disk). When $q_1 = 1$, we simply refer to $q_2$ as $q$ in the text (since in this case the halo potential is axisymmetric). The rotation angles $\alpha$ and $\beta$ specify the orientation of the halo potential with respect to our reference frame. When a disk is included, it is viewed edge-on.}
\label{table: sim_properties}
\end{table}


\subsubsection{Fitting a Stream Track to Simulated Data}\label{sec: fitting_stream_track}
To fit a stream track to our simulation snapshots, we follow a similar approach to \cite{Nibauer}. We first bin particle positions in the $x-y$ plane over a coarse grid to prescribe an initial ordering to stars populating the stream. Bins that are populated with particles are connected to the nearest neighboring bin with a non-zero number of particles. Each bin is connected only once. This defines an initial jagged stream track.  This procedure can fail if the stream intersects itself. In this case, we manually select neighboring bins to fit the elongated axis of the stream. Each particle is then assigned an ordering parameter $\gamma \in [-1,1]$ based on its distance to the initial track. The $\gamma$ parameter is a monotonic phase, encoding progress along the stream track. A differentiable smoothing spline is then fit to the ordered stream, which is used to estimate unit curvature vectors to the track. Additional details of the fitting procedure are discussed in Appendix~\ref{sec: fitting_routine}.

\subsection{Systems Viewed Along a Symmetry Axis}\label{sec: viewed_along_symm}
\subsubsection{Single-Component Model}\label{sec: n_body_results}
\begin{figure*}[htp!]
    \centering
    \includegraphics[scale=0.34]{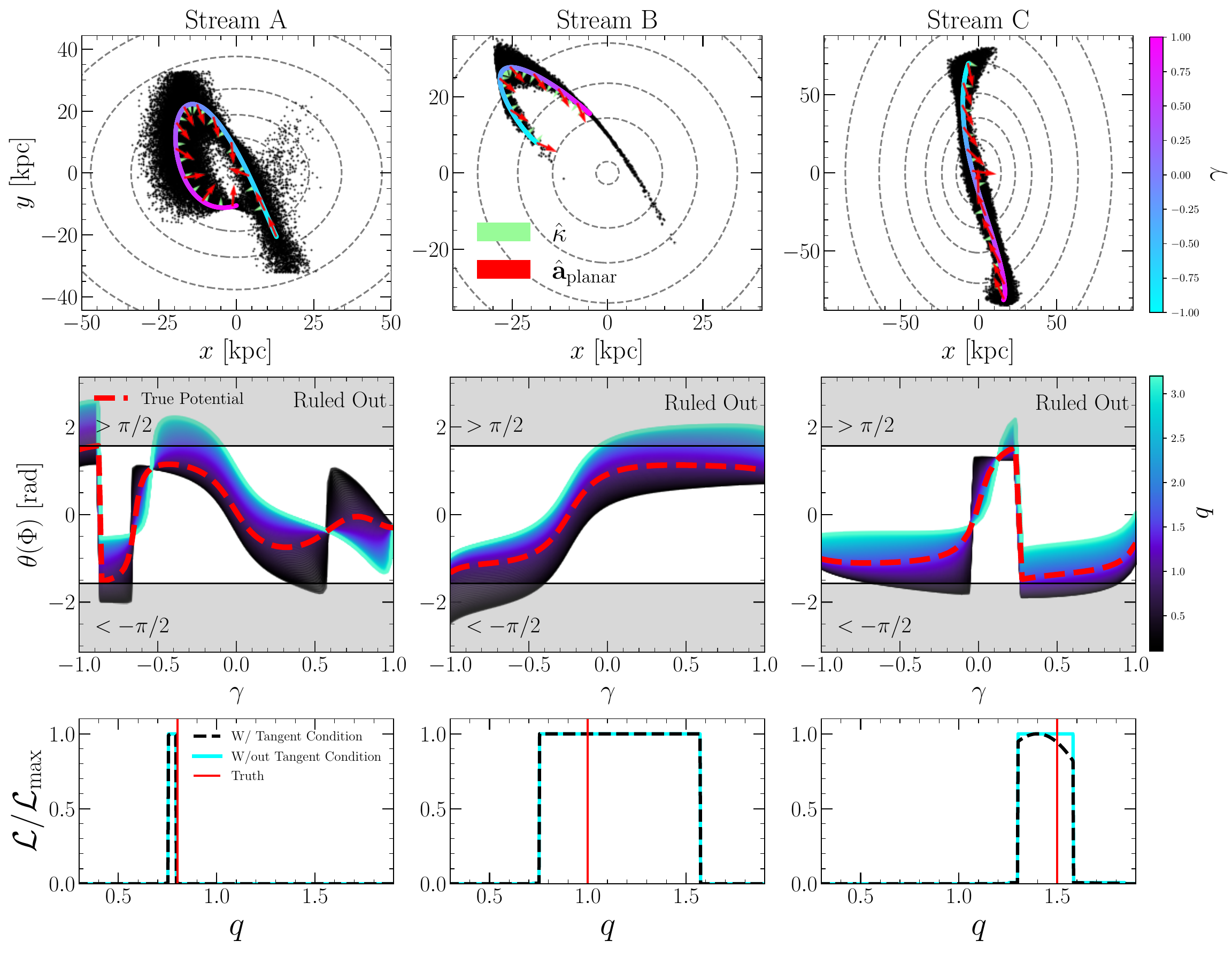}
    \caption{Halo shape inference using 2D projected stream tracks. {\it{Top Row}:} Three streams generated in $N-$body simulations in different external potentials. Dashed gray curves are equipotential contours in the $x-y$ plane. The stream track is shown as the colorful curve in each panel, color-coded by the $\gamma$-value along the track. Planar acceleration unit vectors along the stream are shown as red arrows, and curvature vectors are the green arrows. For the viewing angle considered in this section, $\hat{\boldsymbol{a}}_{\rm planar}$ does not vary with the line of sight distance, $z$. {\it Middle Row:} For each stream (A, B, C), we show the angle $\theta$ between the estimated curvature vector along the track and trial planar accelerations generated from an axisymmetric potential model. The color-code in the middle row corresponds to a trial value for the y-axis flattening parameter, $q$. The gray space indicates ``forbidden regions", where the stream-curvature and planar accelerations are incompatible. {\it Bottom Row:} The likelihood function for each stream with (black dashed curve) and without (cyan curve) the tangent condition implemented. The red line indicates the true flattening parameter.}
    \label{fig: 3Stream_Constraints}
\end{figure*} 
We now apply our stream-curvature analysis to $N$-body streams generated in a ground truth logarithmic potential (Eq.~\ref{eq: log_pot}) with a single free parameter: the $y^\prime$-axis flattening (labeled $q$), and no disk component. The $x^\prime$ flattening ($q_1$ in Eq.~\ref{eq: log_pot}) is set to $1$. We work in a coordinate system that is aligned with the $x^\prime-y^\prime$ plane (i.e., $\alpha = \beta = 0$), so that the halo is viewed along a principal axis. For a fixed point $x_0$ and $y_0$ in this frame, varying $z$ changes only the magnitude of planar accelerations and not their direction. This can be seen by taking equipotential slices along $z$: for the viewing angle considered here, each equipotential slice describes an ellipse with the same axis ratios. Because planar accelerations are perpendicular to each equipotential ellipse, the direction of the planar acceleration vector at a reference point $(x_0, y_0)$ is independent of $z$ for the configuration considered.

In the top row of Fig.~\ref{fig: 3Stream_Constraints}, we show three distinct $N-$body streams generated in gravitational potentials with different flattening parameters, $q = [0.8,1.0,1.5]$ (left to right). Snapshots are taken at $t \approx 10~\rm{Gyr}$. Each column corresponds to a different stream shown in the $x-y$ plane, labeled Stream A, B, and C. Equipotential contours are illustrated as black dashed curves. The spline based stream track is overplotted as the colorful curve. The color-coding of the track corresponds to the $\gamma$-parameter, which encodes a monotonic phase along the stream. The true planar acceleration unit vectors along the track are shown as red arrows, and the curvature vectors are shown as the small green arrows. The latter are estimated from the data using derivatives of the stream track. For stream A, the tail of the stream near $\gamma = -1$ has negligible curvature, so curvature vectors are not shown in this region.

The middle row of Fig.~\ref{fig: 3Stream_Constraints} illustrates the central concept of our analysis: the correct potential model must produce accelerations that fall within $90~\rm{deg}$ of the measured curvature vector (otherwise, the tidal feature would not curve in the observed direction). For a given flattening parameter, $q$, we plot the angle, $\theta$, between the curvature vector and the proposed planar acceleration vector evaluated along the stream track. This yields a family of curves, which are color-coded by the flattening parameter $q$. Flattening parameters that produce $\vert \theta \vert >\pi/2$ anywhere along the stream track are ``ruled out", while those that produce stream-acceleration angles within the white space across the length of the stream are compatible. The $\theta$ versus $\gamma$ profile for the true acceleration field is shown as the dashed red curve.

Streams A and C in Fig.~\ref{fig: 3Stream_Constraints}, middle row, both have a true-potential profile (dashed red) that approaches the gray ruled out regions and abruptly changes. For Stream C this is due to an inflection point, where the curvature vector switches directions by $180~\rm{deg}$ to demarcate a change in the stream's concavity. At an inflection point the planar acceleration vector should then point along the stream track, since $\boldsymbol{a}_\perp = 0$ (\S\ref{sec: zero_curve}). For Stream A, there is a similar inflection behavior in the $\theta$ versus $\gamma$ profile towards the segment of the stream around $\gamma = -1$. However, this segment of the stream is nearly linear, and satisfies the tangent condition $\vert d\phi/d\ell \vert <\epsilon$. Because this portion of the stream has negligible curvature, we treat its curvature vectors as undefined and up-weight accelerations with $\hat{\boldsymbol{a}}_{\rm{planar}} \propto \pm \hat{\boldsymbol{T}}$ in the likelihood (discussed in \S\ref{sec: zero_curve} and \S\ref{sec: likelihood}). Indeed, this can be seen in the top row of Fig.~\ref{fig: 3Stream_Constraints} (Stream A), where the red vectors point along the stream track near $\gamma = -1$. For Stream A, the abrupt change in the $\theta$ versus $\gamma$ profiles near $\gamma = -1$ is simply due to the uncertainty in determining a curvature vector to a nearly linear segment of the stream track.  

The result of our likelihood modeling is illustrated in the bottom row of Fig.~\ref{fig: 3Stream_Constraints} for each stream A-C.  The true input flattening parameter is shown as the red line. For the cyan likelihood curves the tangent condition is not implemented and stream segments with undefined curvature vectors are removed. For the black curves the tangent condition is implemented if $\vert d\phi/d\ell \vert < \epsilon$ anywhere along the stream track. When there is no inflection point or linear stream-segment (e.g. stream B), the two likelihoods are exactly the same. We expect that whether the tangent condition is included or excluded that the resulting likelihoods are consistent, though with the tangent condition providing additional constraining power. This is most clearly seen with Stream C: when the tangent condition is turned on, the likelihood becomes slightly more narrow and provides a more informative constraint on the shape of the potential. For Stream A, only a narrow range of potential models are compatible with the nearly linear segment of the stream towards $\gamma \approx -1$ and the looping segment with $\gamma > -0.5$. In this case including the tangent condition makes a negligible difference to the likelihood, since the full length of the stream already places informative constraints on the shape of the potential. 

Across all three cases, our curvature-acceleration analysis produces accurate constraints on the flattening of the potential with the input parameters supported by the highest likelihood regions. For Stream A in Fig.~\ref{fig: 3Stream_Constraints} the true flattening parameter is at the edge of the highest likelihood region. This is due to the linear stream segment in the top left panel around $(x,y)\approx(15,-15)~\rm{kpc}$; potential models that are too oblate or prolate tilt planar accelerations away from the stream track in this region. This leads to a sharp peak in the likelihood function. Slightly different stream tracks fit using bootstrap resampling of stream stars can shift the highest likelihood region towards the true $q$, motivating a fully probabilistic treatment of the stream track rather than the fixed-track analysis we present here. We discuss the limitations of considering only a single best fit stream track in \S\ref{sec: caveats}. Overall, the strength of the constraints in Fig.~\ref{fig: 3Stream_Constraints} varies depending on the particular stream or potential model considered, and the tangent condition can further limit the range of possible models that are compatible with the stream. Streams with nearly linear segments or inflection points are most useful, since for these systems the direction of the curvature vector changes by $180~\rm{deg}$ over a small area. This discussion is extended in \S\ref{sec: what_makes_informative}, where we highlight which stream morphologies will lead to the most informative constraints on the potential of external galaxies.

\subsubsection{Two-Component Model}\label{sec: disk_halo_misalignment}
\begin{figure*}[htp!]
    \centering
    \includegraphics[scale=0.65]{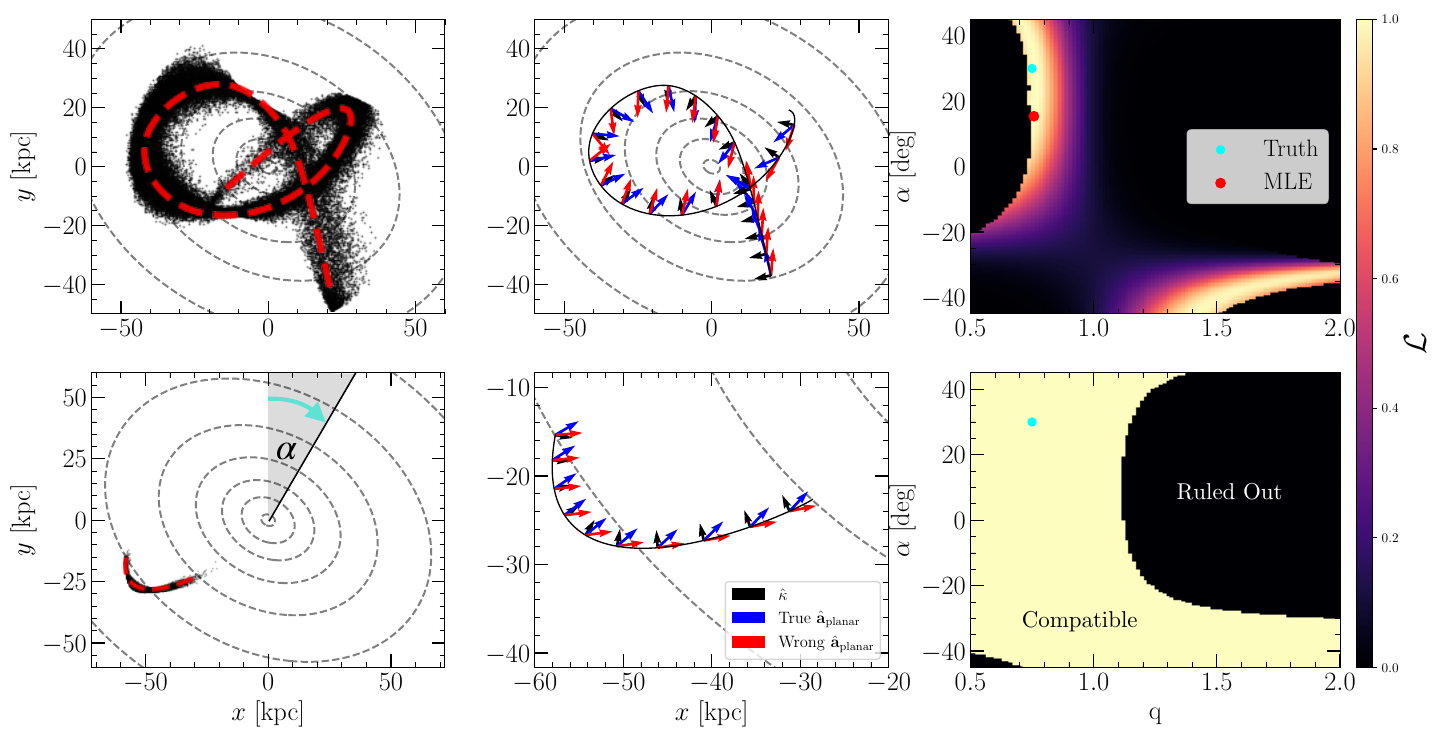}
    \caption{Potential constraints for two streams generated in the same background potential, consisting of a Miyamoto-Nagai disk and an axisymmetric halo.
    The disk is edge-on at $y=0$, and the halo has flattening parameter $q = 0.75$. The halo flattening axis is rotated by an angle $\alpha = 30~\rm{deg}$ with respect to the disk. Equipotential contours are plotted at $z = 0$. Stream tracks for both systems are shown in the left column, and isolated in the middle column. Black vectors indicate unit curvature vectors to the track, while blue and red vectors indicate trial planar accelerations generated from the correct potential model
   and the incorrect potential model, respectively. The right column illustrates the likelihood surface for both tidal features. Both the flattening direction ($\alpha$) and flattening ($q$) are free parameters. The input parameters are shown as the cyan point, and the maximum likelihood estimate (MLE) is the red point. The stream in the top row has evolved for $10.5~\rm{Gyr}$, while the bottom row corresponds to $1.5~\rm{Gyr}$. Longer, and therefore older streams typically produce more informative constraints on the potential.}
    \label{fig: Rotated_Potential}
\end{figure*}

So far, we have only considered a single component potential model; however, we expect galaxies to have a stellar and a dark matter component. In this section we explore the possibility of detecting both components, and their relative orientation.

Cosmological simulations predict that dark matter halos can be aspherical and tilted with respect to the disk \citep{2012JCAP...05..030S, 2019MNRAS.484..476C,2021ApJ...918....7E, 2022arXiv221208880P}. Indeed, in the Milky Way there is observational evidence for a disk-halo misalignment with an ellipsoidal dark matter halo that is tilted with respect to the galactic plane \citep{2009ApJ...703L..67L,2022AJ....164..249H, 2022arXiv221014983P}. Motivated by these findings, we now treat the disk-halo misalignment angle as a free parameter, and consider constraining its value from projected stream morphology.

The simulations utilized in this section are outlined in \S\ref{sec: gen_Nbody}. Importantly, we rotate the flattening direction of the axisymmetric halo by an angle $\alpha = 30~\rm{deg}$ relative to a perpendicular axis passing through the midplane of the disk. We refer to this angle as the disk-halo angle. A $30~\rm{deg}$ misalignment is similar to recent estimates of disk-halo misalignment in the Milky Way \citep{2022AJ....164..249H}. The disk-halo angle, $\alpha$, in our frame is illustrated in the bottom left panel of Fig.~\ref{fig: Rotated_Potential}. In this frame the disk is viewed edge-on, with the $y$-axis perpendicular to the midplane. 

In the combined potential of the disk and rotated halo, we simulate two streams in exactly the same external potential. The initial conditions for these simulations were generated using the random sampling procedure described in \S\ref{sec: gen_Nbody}. Snapshots of both $N-$body realizations are shown in the leftmost-column of Fig.~\ref{fig: Rotated_Potential}. We purposely apply our analysis to two streams at very different stages in their evolution, in order to demonstrate how our constraints in this section depend on when the stream is observed throughout its evolution. The stream in the top row of Fig.~\ref{fig: Rotated_Potential} corresponds to a $t \approx 10.5~\rm{Gyr}$ snapshot, and $t \approx 1.5~\rm{Gyr}$ for the bottom row.   

For both $N-$body systems in Fig.~\ref{fig: Rotated_Potential} the stream track is illustrated with the dashed red curve. For the stream in the top row, the tangent condition is satisfied along the bottom right segment of the track around $(x,y)\approx(20,-30)~\rm{kpc}$. The shorter stream in the bottom row has no change in concavity nor a locally linear segment. 

In the middle column of Fig.~\ref{fig: Rotated_Potential} the curvature vectors (black vectors) for each tidal feature are shown, along with trial accelerations in the $x-y$ plane (blue and red vectors). The bottom middle row is zoomed-in around the smaller stream for clarity. The blue vectors correspond to the correct potential model, which has the halo component misaligned with the disk. The red vectors are generated from an incorrect potential with no disk-halo misalignment. Indeed, for both tidal streams there are several instances where the red vectors (incorrect model) subtend an angle greater than $90~\rm{deg}$ from the estimated curvature vector. However, under the correct potential model with the halo oriented at an angle relative to the disk (blue vectors), the planar accelerations and estimated curvature vectors are compatible.

When constraining the shape of the potential, we set the mass ratio between the disk and halo component to the true value (where the enclosed mass within 40~\rm{kpc} is $M_{\rm Disk} \approx 1.2\times10^{10}M_{\rm Sun}$ and $M_{\rm Halo} \approx 3.8\times10^{12}M_{\rm Sun}$). In \S\ref{sec: mass_ratios}, we discuss constraining the mass ratio as a free parameter.  For the flattening parameter we sample $q$ in the range $[0.5, 2]$. For the disk-halo angle, $\alpha$, note that unit vector accelerations in the $x-y$ plane are equivalent when the flattening axis is rotated by $90~\rm{deg}$ with $q \xrightarrow{} 1/q$.  To reduce degeneracies when evaluating the likelihood, rotations are sampled in the interval of $\alpha \in [-45,45]$ degrees. This approach still results in some degeneracies with rotation, as an oblate halo with $\alpha = 40~\rm{deg}$ degrees is not significantly different from a prolate halo with $\alpha = -40~\rm{deg}$, even though the two models are not exactly the same. Constraints for the flattening parameter, $q$, and disk-halo angle, $\alpha$, are shown in the rightmost column of Fig.~\ref{fig: Rotated_Potential}. In the top right panel, the maximum likelihood estimate (MLE) is shown as the red point while the true parameters are the cyan point in both panels (top and bottom right).

For the case of the extended stream in the top row of Fig.~\ref{fig: Rotated_Potential}, a spherical halo is ruled out at the $\sim 2\sigma$ level using a likelihood ratio test between the best fit model and a spherical halo model with $q=1$. A non-zero disk-halo angle is slightly preferred, though the significance of the misalignment is below the $1\sigma$ level. However, the constraints imply that equipotential contours are flattened when measured along an axis with $\alpha = 30~\rm{deg}$, or stretched for $\alpha = -30~\rm{deg}$. As discussed above, these two solutions are complementary, since an oblate logarithmic potential can be approximated as a rotated prolate potential in Eq.~\ref{eq: log_pot}. From the MLE for this stream (red point; Fig.~\ref{fig: Rotated_Potential}), we find that there is a slight preference for the $q < 1$ mode over the $q > 1$ mode.

For the less extended stream in the bottom left, the likelihood surface is split into two regions: one where planar accelerations are compatible with the estimated curvature vectors (yellow; high likelihood) and one where they are not (black; low likelihood). Because the stream has a much smaller length compared to the extended stream in the top left, an incorrect potential model can produce incompatible planar accelerations for a large number of evaluation points along the stream (i.e., nearly $1/3$ to $1/2$ of the evaluation points have incompatible planar accelerations when sampling in the black regions of the likelihood surface). This leads to the large contrast between high and low (near zero) probability regions in the likelihood panel. Because there is no inflection point nor linear segment along the stream, the tangent condition (\S\ref{sec: zero_curve}) does not provide any additional improvement to the constraints.

While the constraints in the bottom right panel of Fig.~\ref{fig: Rotated_Potential} are far less informative than in the top right panel, we can still make interesting statements about the shape of the halo potential as a function of its relative orientation to the disk. For instance, if we assume that the disk and halo are aligned (i.e., $\alpha = 0$) then we find $q \lesssim 1.1$. For $\alpha = -40~\rm{deg}$ all flattening parameters in the sampled interval are compatible. In this case, incorporating more streams in the constraints can break degeneracies, though it is still possible to consider a meaningful range of halo geometries (i.e. $q \lesssim 1.1$) as a function of disk-halo misalignment angles.  

The results presented in this section demonstrate that while we can place tight constraints on the flattening of the dark matter halo (Fig.~\ref{fig: Rotated_Potential}, top row), the disk-halo misalignment angle is more difficult to constrain from projected stream morphology. More extended streams with multiple loops or an inflection point represent cases where both parameters can be determined at a higher confidence level.  
 
\subsubsection{Limits on Mass Ratios}\label{sec: mass_ratios}
In the absence of dynamical information (e.g., velocity of the stream), the mass enclosed by the tidal feature is degenerate with viewing angles, distance gradients, and the 3D velocity distribution of stars populating the tidal feature (see, e.g., \citealt{2013MNRAS.435..378S, 2022ApJ...941...19P}). This is clear from Eq.~\ref{eq: a_perp_over_mag}: adjusting the total mass of the system by scaling the amplitude of the overall potential does not alter the output of the $\sgn$ function. This makes intuitive sense: unit vector accelerations do not change if the total mass of the system increases or decreases uniformly (that is, for a potential function of the form $\Phi(\boldsymbol{x}) \equiv GMf(\boldsymbol{x}) \xrightarrow{} c GMf(\boldsymbol{x})$ for some scalar $c$, unit vector accelerations are invariant to this scaling: $\hat{\boldsymbol{a}} \xrightarrow[]{} \hat{\boldsymbol{a}}$). However, for two-component potentials (e.g., a disk and halo) of the form 
\begin{equation}
\begin{split}
    \Phi(\boldsymbol{x}) &= m_1 f_1(\boldsymbol{x}) + m_2 f_2(\boldsymbol{x}) \\
    &= m_1\left( f_1(\boldsymbol{x}) + \frac{m_2}{m_1}f_2(\boldsymbol{x})\right),
\end{split}
\end{equation}
 the ratio $m_2/m_1$ can be constrained, since $\hat{\boldsymbol{a}}$ depends on the ratio $m_2/m_1$. The amplitudes $m_1$ and $m_2$ are related to the characteristic mass of each potential component, with the dimensions of (e.g.) $GM$ (so that $f_1$ and $f_2$ have dimensions of inverse length). Varying the mass ratio between potential components will change the direction of unit vector accelerations, as one component ``pulls" unit accelerations away from the other. This is especially relevant for combined potentials with
an anisotropic term (e.g., a disk or flattened halo), since aspherical mass distributions can significantly alter the direction of unit vector accelerations in the plane. Regardless of the geometry of the mass distribution, the correct mass ratio yields an overall potential model with planar accelerations that are consistent with the projected curvature vectors along the stream. 

This provides a means to estimate a mass ratio between a halo component and a disk component, even if the mass of both components are individually unknown. If the baryonic mass of the disk is determined independently, e.g., using mass-to-light ratios or other scaling relations, then the estimated ratio $m_2/m_1$ can be used to solve for the mass of the dark matter component interior to the stream. If sufficiently precise, this can provide a ``mass-ladder" for a single galaxy, where a stellar mass estimate can be used to calibrate the halo mass estimate derived from stream curvature. 

\begin{figure}[tp!]
    \centering
    \includegraphics[scale=0.33]{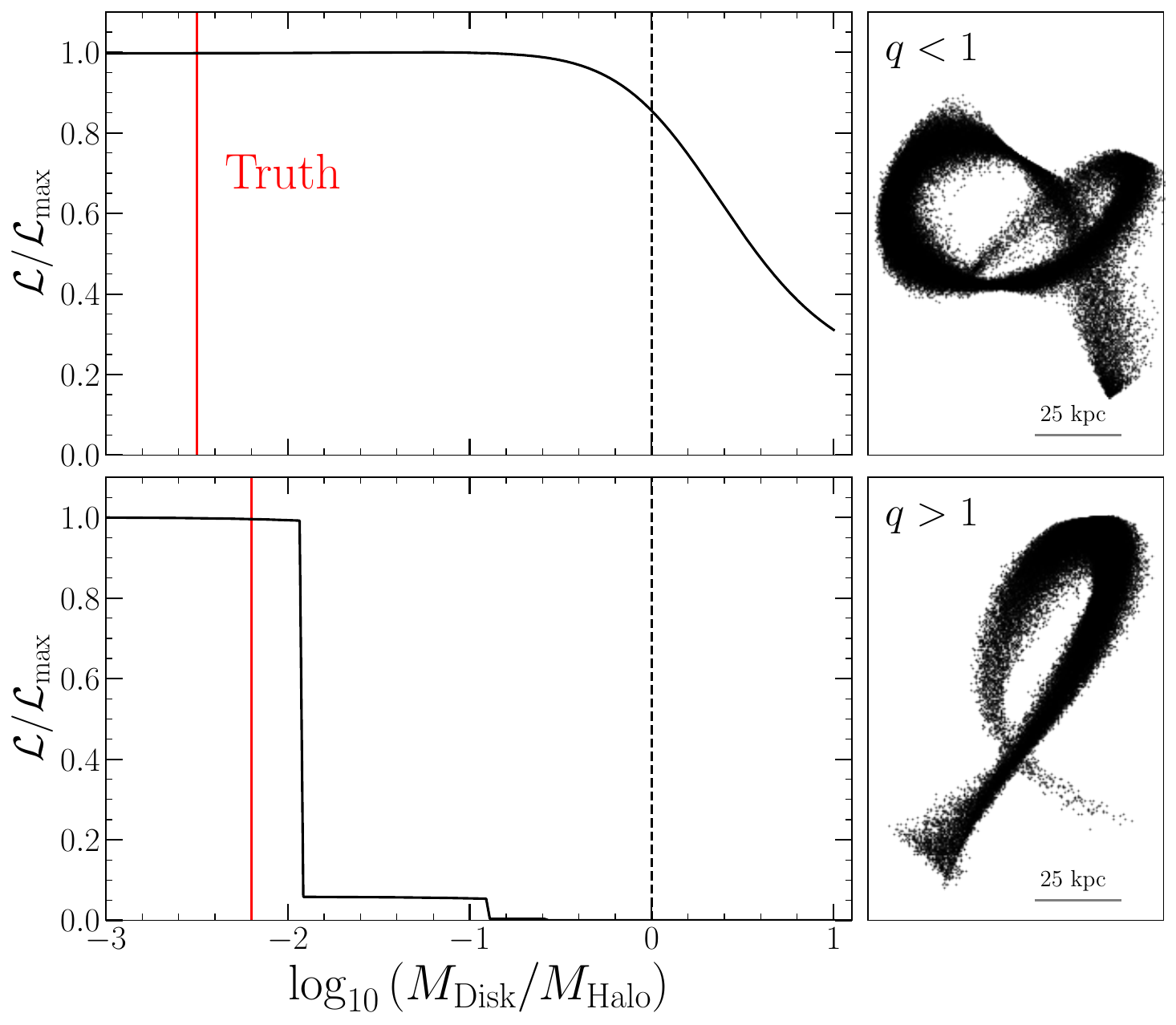}
    \caption{Inferring the disk-to-halo mass ratio from 2D projected stream tracks. {\it Top Panel:} Likelihood function for the stream in Fig.~\ref{fig: Rotated_Potential}, top row, generated in an oblate halo potential. The stream is also reproduced in the top right panel. At the best fit flattening ($q$) and disk-halo angle ($\alpha$), we evaluate the likelihood as a function of disk-to-halo mass ratios (defined as the enclosed mass ratio within a radius of $40~\rm{kpc}$). The true input mass ratio is shown as the red line. {\it Bottom Panel:} Same as the top row, but for a stream generated in a prolate halo potential with $q = 1.25$ and a slightly more massive edge-on disk. In both cases we infer the presence of a dark matter halo more massive than the disk (black dashed line).}
    \label{fig: mass}
\end{figure} 

We demonstrate the ability to place limits on mass ratios using the $N-$body stream in Fig.~\ref{fig: Rotated_Potential}, top row. We fix $q$ and the disk-halo angle to the best fit values from Fig.~\ref{fig: Rotated_Potential} (shown as the red scatter point). We constrain the mass ratio $
\log_{10}\left(M_{\rm Disk}/M_{\rm Halo}\right)$ for the two-component potential model used in \S\ref{sec: disk_halo_misalignment}. In this section $M_{\rm Disk} / M_{\rm Halo}$ refers to the mass enclosed ratio within a radius of $40~\rm{kpc}$, since this is roughly the extent of the stream in the $x-y$ plane. The scale length and height of the disk component are set to the true values used in our simulations, under the assumption that these can be roughly determined from the photometric properties of the disk. The core radius of the logarithmic potential component is also set to the true value used in simulations, though reasonable variations on the core radius do not alter our conclusions. 

At the best fit $q$ and disk-halo angle ($\alpha$), Fig.~\ref{fig: mass} (top panel) illustrates the model likelihood for a range of mass ratios. Using the true values for these parameters does not alter our discussion, since the MLE for $q$ and $\alpha$ is so close to the input parameters (both in likelihood and distance to the truth). The true mass ratio is illustrated with the red line. Mass ratios to the right of the dashed black line have $M_{\rm Disk} > M_{\rm Halo}$. The likelihood sampled over mass ratios plateaus to the left of the black line, implying that the data (slightly) prefer a two-component mass model with a dominant halo component over the disk component. A likelihood ratio test shows that a two component model is preferred at the $1.5\sigma$ level, compared to a model with no dark matter halo or a dominant disk component. Note that we may only place an upper limit on $M_{\rm Disk}/M_{\rm halo}$ from Fig.~\ref{fig: mass}.  

A limiting factor in constraining the disk-to-halo mass ratio is the degree of similarity between the acceleration field of the two mass components. For instance, the 
topology of the unit vector acceleration field produced by an oblate halo potential is similar to that of a disk. In contrast, the unit vector acceleration field produced by a prolate halo potential is distinct from that of a disk. In the latter case, the direction of unit vector accelerations is more sensitive to the relative mass amplitude between the two components, so we expect our curvature-based analysis to provide more stringent limits. We demonstrate this point in the bottom panel of Fig.~\ref{fig: mass}, where we apply the same mass ratio analysis to a stream generated in a prolate logarithmic potential ($q = 1.25$), with a major axis perpendicular to the disk's midplane. The true flattening parameter is used in our inference, and only the mass ratio is allowed to vary. A slightly larger disk mass is also used compared to the top panel, though our conclusions do not change for small variations of the disk mass. For the prolate halo potential, the likelihood function implies a strict upper bound of $\log_{10}\left(M_{\rm Disk}/M_{\rm Halo}\right) \approx -1.95$, corresponding to the mass enclosed ratio within $40~\rm{kpc}$. This is a much more informative limit in comparison to the top panel of Fig.~\ref{fig: mass}, where the degree of similarity between the oblate halo potential and the disk limits what can be inferred about the relative mass of the two components using only stream curvature. Still, in both cases the input mass ratio (red line) is recovered within the highest likelihood region.

The results presented in this section suggest that that extragalactic streams in halo potentials with  significantly different shape parameters from the disk (i.e., prolate, or misaligned rather than strongly oblate) provide strong prospects for placing limits on disk-to-halo mass ratios. In future work we intend to explore a wider range of potential models and configurations, to determine the viability of constraining mass ratios from projected stream morphology. 

\subsection{General Viewing Angle}\label{sec: 3D_shape}
\begin{figure*}[htp!]
    \centering
    \includegraphics[scale=0.6]{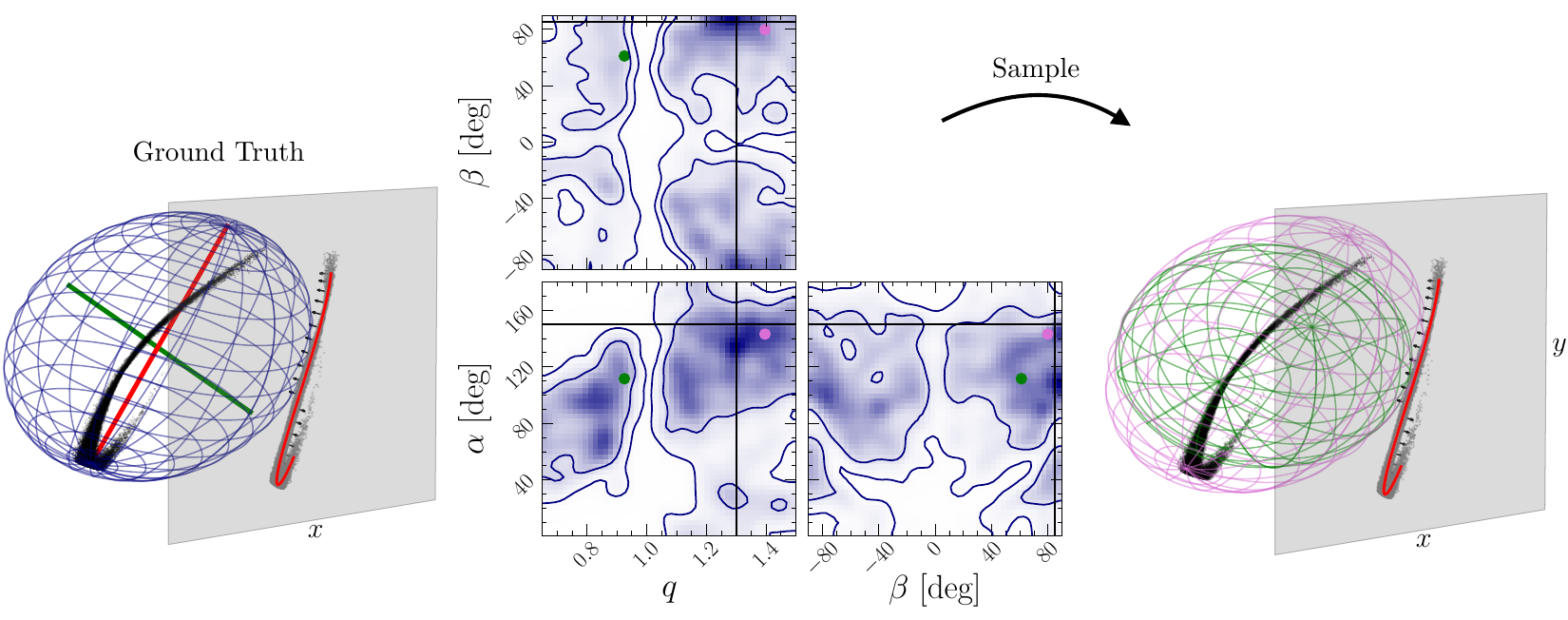}
    \caption{Constraining the halo shape observed at a general orientation, i.e., not along the symmetry axis. {\it Left:} 3D spatial view of a stellar stream, generated in an $N$-body simulation with an external logarithmic potential with flattening parameter $q = 1.3$. The blue surface is an equipotential ellipsoid. The system has been rotated by spherical polar angles $\alpha$ and $\beta$. The major axis is shown as the red line. The stream is projected onto the plane of the sky ($x,y$ plane), illustrated by the gray plane. A track is fit to the projected stream in red, and a few unit curvature vectors are shown along the track.  Middle: Constraints on the 3D shape of the potential from the projected stream track (red curve in the left panel). Input parameters are shown as black lines. Regions of maximal likelihood are close to the ground truth. Right: Two samples from the posterior are illustrated as equipotential ellipsoids. Pink and green ellipsoids correspond to the pink and green points in the middle panel. }
    \label{fig: 3DStream_Plot}
\end{figure*} 
In the preceding sections we assumed that the observed galaxy is viewed along one of its principal axes (i.e., perpendicular to the flattening axis of the halo potential), so that the inferred flattening, disk-halo angle, and disk-to-halo mass ratio can be interpreted as global constraints. If this assumption is broken, then the best-fit flattening parameter, $q$, characterizes the axis ratios of the best fit projected equipotential slice taken over the length of the stream. This does not need to coincide with the global shape of the potential, but characterizes a cross section of the full equipotential surface. For example, slices of an oblate (prolate) ellipsoid are oblate (prolate), or circular at worst. We have verified that the best fit flattening, $q$, for an inclined oblate or prolate potential continues to favor oblate and prolate flattening parameters, respectively.

Importantly, the best fit equipotential slice is specialized to the distance of the stream. For a highly elongated or flattened potential that is inclined with respect to the line-of-sight, distinct portions of the stream can prefer different flattening parameters, depending on the magnitude of the distance gradient along the stream. The likelihood devised in \S\ref{sec: likelihood} accounts for this possibility by sampling acceleration vectors over the line of sight, allowing us to---in principle---constrain the 3D shape of the potential by stitching together the range of elliptical slices that each segment of the stream is compatible with. We now explore to what extent the 3D shape of the potential can be recovered from the projected track of a tidal stream. We consider an axisymmetric halo potential in \S\ref{sec: axisymm_general}, and a triaxial halo potential in \S\ref{sec: triaxial_potential}.

\subsubsection{Axisymmetric Potential}\label{sec: axisymm_general}
We assume the same logarithmic potential (Eq.~\ref{eq: log_pot}) as in the previous sections with $q_1 = 1$ and $q_2$ free (simply referred to as $q$). We also attempt to constrain the rotation parameters $(\alpha, \beta)$ introduced in \S\ref{sec: observing_sims}. These parameters characterize the 3D orientation of the dark matter halo relative to our reference frame (Eq.~\ref{eq: lin_transf}-\ref{eq: rot_mat}). 

The model parameters are $\boldsymbol{m} = \left(q,\alpha,\beta\right)$. We sample $\alpha \in [0,\pi]$, and $\beta \in [-\pi/2,\pi/2]$. The limits in $\beta$ do not actually restrict the range of allowable geometries, since the logarithmic halo potential repeats orientations outside of this range.

As a test case we use a snapshot ($t \approx 10~\rm{Gyr}$) of a stream generated in the axisymmetric logarithmic halo potential  with flattening parameter $q = 1.3$. We rotate the system by the angles $(\alpha, \beta) = (150^\circ, 85^\circ)$, so that equipotential ellipsoids are ``tilted" significantly towards the line-of-sight, and rotated slightly away. This orientation represents a scenario where our previous assumption of viewing the halo along a principal axis is broken. The rotated system is illustrated in the left panel of Fig.~\ref{fig: 3DStream_Plot}, where we plot the 3D stream in black and an equipotential ellipsoid in blue. The plane of the sky is depicted as the gray slate. The ``observed" stream projection is shown as the gray points. Our fit to the projected stream track is illustrated by the red curve, and unit curvature vectors to this track are plotted in black for a few locations. As before, the red track and its derivatives in the $(x,y)$ plane are the only inputs to the analysis. When sampling accelerations over the line of sight, we consider a prior distance interval within $\pm 100~\rm{kpc}$ (see \S\ref{sec: distance_gradients} for a discussion of distance gradients).

In this section posterior samples are obtained using the dynamic nested sampling package, \texttt{dynesty} \citep{2020MNRAS.493.3132S}. Constraints on the 3D orientation and flattening of the projected stream are shown in the middle set of panels in Fig.~\ref{fig: 3DStream_Plot}. The true parameters are the black lines. Each contour (68 and 95\% regions) is shaded according to the density of samples it contains. The islands of high probability within each $68\%$ region comes from the nearly linear segment of the projected stream, where the tangent condition (\S\ref{sec: zero_curve}) is satisfied.
 There are several modes and high probability islands due to degeneracies in reconstructing the 3D shape of the potential from only the 2D morphology of the stream. For instance, a prolate ($q>1$) potential can also be approximated by a rotated oblate ($q<1$) potential. This is the source of the two modes in flattening: $q<1$ and $q>1$. Placing tighter priors on $\alpha$ and $\beta$ can significantly alleviate degeneracies (e.g., from measurements of the disk inclination), though in the present work we sample over all possible orientations. Note however that a spherical halo with $q = 1$ is disfavored in this example. While the posterior distribution over these parameters is degenerate and multimodal, we can rule out a halo component with a flattening axis perpendicular to the line-of-sight ($\alpha = \beta = 0$) at the $\sim 2\sigma$ level. Furthermore, the highest posterior density regions are indeed close to the ground truth logarithmic potential, both in its global flattening and 3D orientation.

\begin{figure*}[tp!]
    \centering
    \includegraphics[scale=0.65]{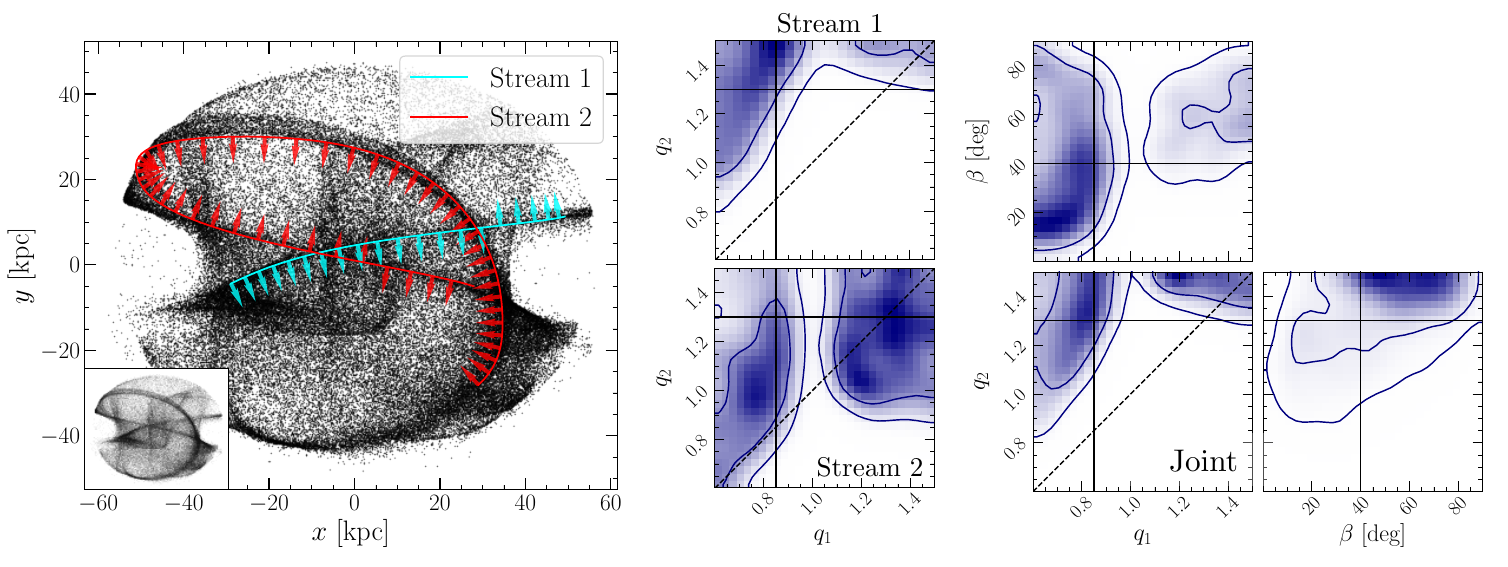}
    \caption{ Limits on the axis ratios and orientation of a triaxial potential from 2D projected stream tracks. {\it Left}: Simulation snapshot of tidal features generated in a triaxial logarithmic potential. At $\sim 8.8~\rm{Gyr}$ several streams and shells have formed. Tracks are fit to two visually obvious streams (Stream 1 in cyan, Stream 2 in red). The curvature vectors to each track is shown by the red and cyan arrows. The bottom left inset shows the same snapshot without any superimposed stream tracks. {\it Middle:} Using the curvature vectors along the two stream tracks, we sample over the $x^\prime$ and $y^\prime$ flattening parameters $(q_1,q_2)$ and rotation angle $\beta$ to identify what combination of shape parameters could have produced the observed curvature. Marginal constraints in the $(q_1, q_2)$ plane are shown for each stream independently. Input parameters are the solid lines, and the dashed lines trace $q_1 = q_2$.
    {\it Right:} Posterior constraints on the triaxial potential derived from both streams. The true parameters are the solid lines. A spherical halo with $q_1 = q_2=1$ is ruled out, and the correct input parameters are recovered within the 68\% region.  }
    \label{fig: Triaxial_Constraints}
\end{figure*} 

In the right panel of Fig~\ref{fig: 3DStream_Plot} we plot equipotential ellipsoids corresponding to two samples from the posterior distribution in $(q,\alpha,\beta)$ shown as the scatter points in the middle panel (green and pink points correspond to the green and pink ellipsoids, respectively). Posterior draws with $q<1$---while less likely than those with $q > 1$---tend to be rotated to have prolate elliptical level sets when sliced along the line of sight. This is the expected behavior, since planar accelerations are orthogonal to line-of-sight slices of equipotential ellipsoids, and our analysis constrains a potential model through its planar acceleration field. 

Still, there are significant uncertainties and degeneracies in reconstructing the 3D shape of the potential from only the projected track of a single stream. As we will discuss in \S\ref{sec: what_makes_informative}, certain stream morphologies will be most useful for constraining the potential. Incorporating multiple streams in the 3D analysis presented here is also computationally feasible, and can break degeneracies to provide more stringent constraints. Perhaps surprisingly, our results show that the continuous nature of streams along the line of sight enables us to utilize projected stream curvature to estimate 3D properties of the observed galaxy's halo, even in the complete absence of line-of-sight information. 

\subsubsection{Triaxial Potential}\label{sec: triaxial_potential}
We now explore constraining the axis ratios of a triaxial potential from the projected shape of tidal streams. The analysis works the same as before: by measuring the curvature to a stream, we can identity a range of potential models that are compatible with the observed curvature. We generate tidal features in a triaxial version of the logarithmic potential defined in Eq.~\ref{eq: log_pot}. Mock-observations are produced by rotating the stream and potential by the angles $\alpha = 0~\rm{deg}$ and $\beta = 40~\rm{deg}$.
The simulation procedure is further discussed in \S\ref{sec: gen_Nbody}.

A snapshot of the rotated system in the plane of the sky $(x,y)$ is shown in Fig.~\ref{fig: Triaxial_Constraints} (left panel), corresponding to $t = 8.85~\rm{Gyr}$. While we evolve a single satellite, several streams and shells form, each delineating its own family of orbits (this is expected for a triaxial model; see \citealt{2022arXiv221211006Y}). We use splines to fit tracks to two visually obvious, elongated streams (labeled Stream 1 and 2) after manually selecting a few representative points to construct the track. Unit curvature vectors to the two tracks are shown. For both streams there is a change in concavity. 

We treat each stream independently in the likelihood, and constrain the two flattening parameters $(q_1,q_2)$ and the rotation angle $\beta$. When sampling acceleration vectors from the triaxial potential, we fix $r_c$ to the true value (16~\rm{kpc}), and the choice of $v_c$ is arbitrary since we only constrain the direction of acceleration vectors and not their magnitude. Acceleration vectors are sampled along the line of sight within $\pm 70~\rm{kpc}$ of the host galaxy, which is roughly the extent of the tidal debris in the $x-y$ plane. The posterior distribution over the model parameters are again drawn using \texttt{dynesty} \citep{2020MNRAS.493.3132S}. 

We present constraints in Fig.~\ref{fig: Triaxial_Constraints} derived from each stream separately (middle column), and jointly (right set of panels). The combined constraints are produced by sampling from a joint posterior, which is constructed by taking the sum of the total log-likelihoods for each independent stream. The constraints in the middle column are marginalized over the rotation parameter $\beta$. Contours correspond to regions of 68 and 95\% confidence. The true parameters are shown as the solid lines, and dashed lines correspond to $q_1 = q_2$ (for an axisymmetric potential). A key advantage of our method is that we are immediately able to identify which aspects of a stream inform constraints on the potential (e.g., \S\ref{sec: n_body_results} and the $\theta(\Phi)$ v.s. $\gamma$ plots). For Stream 1, most information comes from the nearly linear segment of the track around $x = 34~\rm{kpc}$ in the left panel of Fig.~\ref{fig: Triaxial_Constraints}. For streams with nearly linear segments or inflection points, we expect planar accelerations to align with their track (\S\ref{sec: zero_curve}). For stream 2, the inflection point around $x = 5~\rm{kpc}$ again dominates the constraints, though the extended looping segment helps place limits on the orientation of the flattening axes. 

Perhaps surprisingly, Stream 1 appears to provide more information about the potential compared to Stream 2. This is because Stream 1 has a flat segment around $x = 34~\rm{kpc}$, activating the tangent condition for several evaluation points. Stream 2 has an inflection point, but only satisfies the tangent condition at a single evaluation point (the inflection point). Streams with extended flat regions are especially useful for constraining the potential, since 
flat segments have a vanishing acceleration component. Furthermore, the change in concavity around the flat segment of Stream 1 also limits the range of possible orientations for the potential.

From the joint constraints, a spherical halo is ruled out ($>2\sigma$) due to the presence of inflection points, and a triaxial halo with $q_2 > q_1$ is preferred. The correct input parameters are recovered within the 68\% region. When combining information from both streams, we find only a marginal improvement in constraining power compared to the constraints from Stream 1 alone. However, there is a region at low $q_1$ in the joint constraints that receives less support due to the inclusion of Stream 2. 

Unlike generative simulation based methods, including additional streams in our analysis is not computationally expensive, since we simply compare acceleration vectors to curvature vectors along each stream track. In principle, constraining the potential from, e.g., tens of streams simultaneously would be practical, provided that a differentiable track is fit to each stream. Future observations are expected to reveal a substantial population of streams in M31 (i.e., \citealt{2022ApJ...926..166P}): this work provides an efficient method to constrain the potential from an ensemble of tidal features.

We use this section to demonstrate that for galaxies with multiple streams or streams with constraining features (i.e., inflection points, flat segments, saddle points), it can be feasible to constrain shape parameters of the halo beyond axisymmetry. A more realistic potential model may also include a disk component, and parameters characterizing the relative orientation of the triaxial halo to the disk. In the absence of velocity information, constraints will typically be degenerate, though can still provide an indication of the range of halo geometries that are compatible with stream curvature.

\section{Measuring the Gravitational Potential of NGC 5907}\label{sec: application_ngc59076}

\begin{figure}
    \centering
    \includegraphics[scale=0.65]{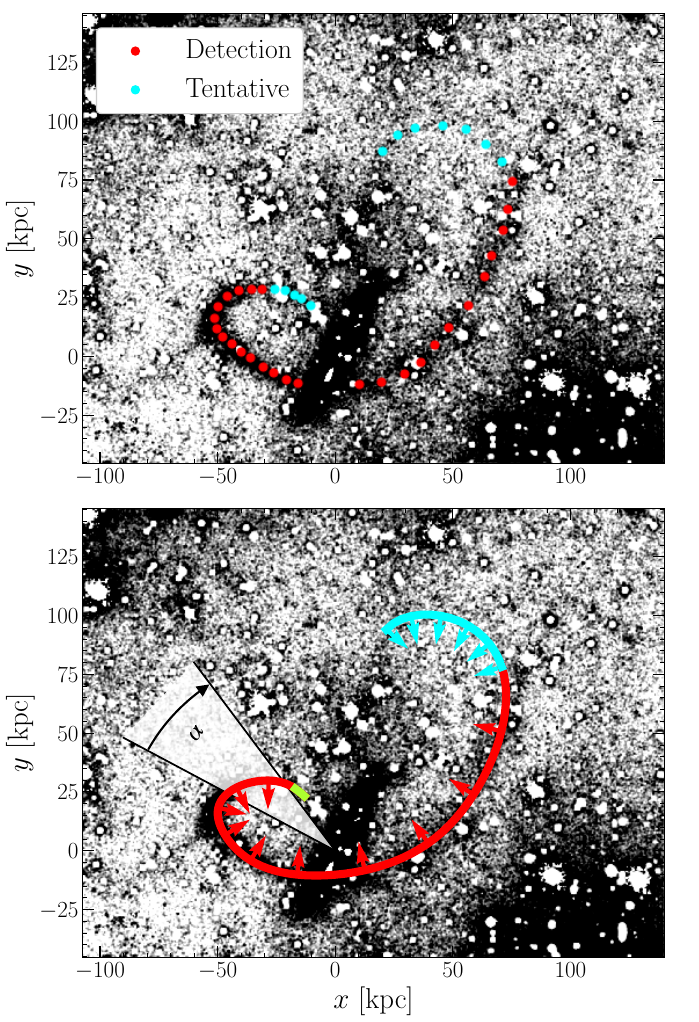}
    \caption{ Dragonfly imaging of NGC 5907 and its stellar stream \citep{2019ApJ...883L..32V}. In the top panel, scatter points indicate regions of the stream that are detected at high confidence (red) and the fainter tentative regions of the stream (cyan). In the bottom panel, we show the spline-based fit to the stream, along with its unit curvature vectors. The green segment of the stream-track satisfies the tangent condition, $\vert d\phi/d\ell \vert < \epsilon$. We illustrate  the disk-halo angle, $\alpha$, in the bottom panel. In \S\ref{sec: two_comp}, we treat $\alpha$ as a free parameter. }
    \label{fig: ngc5907}
\end{figure} 

We now apply the method introduced in this work to low surface brightness imaging of the tidal feature surrounding the galaxy NGC 5907. NGC 5907 is an edge-on spiral galaxy at a distance of $\sim$17~\rm{Mpc} \citep{2016AJ....152...50T}, with stellar mass $\sim 8 \times 10^{10} M_{\rm Sun}$ \citep{2016AJ....152...72L}. The detection of a stream associated with the galaxy is first discussed in \citet{1998ApJ...504L..23S} and \citet{1999AJ....117.2757Z}. Subsequent imaging of the galaxy and its tidal stream revealed a tentative second loop \citep{2008ApJ...689..184M}, which was reproduced in the context of $N-$body simulations. Recent imaging from the Dragonfly Telephoto array finds evidence for a single curved stream, with an angular length $\approx 45^\prime$ \citep{2019ApJ...883L..32V}. Further evidence for a single curved stream consistent with \citet{2019ApJ...883L..32V} is presented in \citet{2019A&A...632L..13M}. 

\begin{figure*}
    \centering
    \includegraphics[scale=0.65]{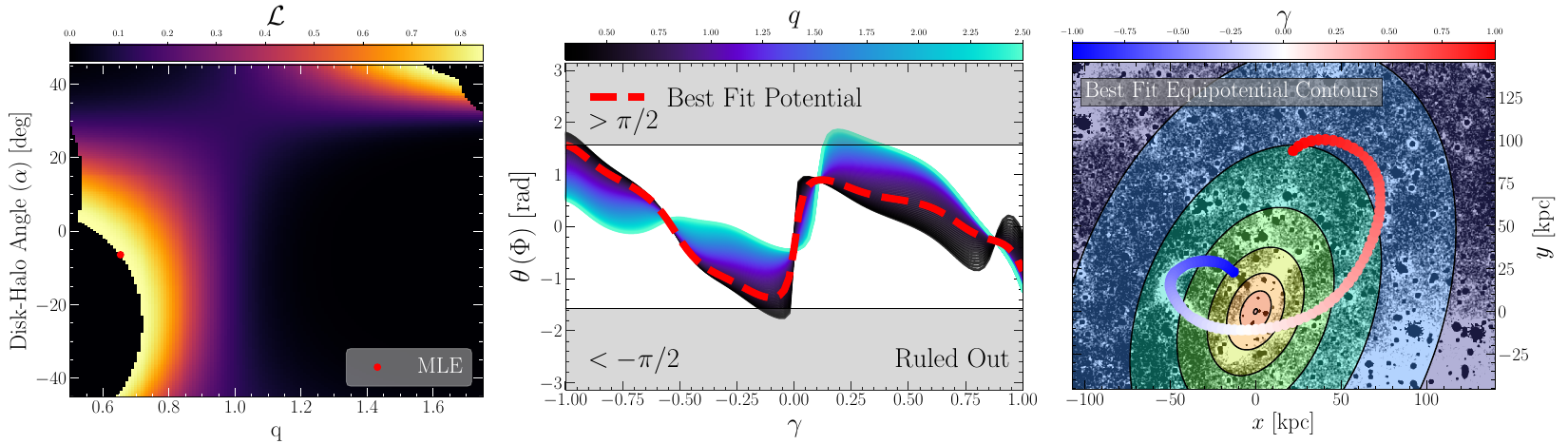}
    \caption{Constraints on the gravitational potential of NGC 5907 for a single component model. {\it Left:} Using the curvature vectors to the stream track surrounding NGC 5907 (Fig.~\ref{fig: ngc5907}), we fit a single-component logarithmic halo potential with flattening parameter $q$. We allow both the flattening axis and parameter to vary. The disk-halo angle ($\alpha$) is measured relative to an axis perpendicular to the disk. The likelihood surface for this two-parameter model is illustrated. The MLE is the red point. {\it Middle:} At the best fit disk-halo angle ($\alpha \approx -7^\circ$), we plot the stream-acceleration angle $\theta$ versus the monotonic phase-parameter $\gamma$. Each $\theta$ versus $\gamma$ profile is color-coded by a trial flattening parameter, $q$. The $\theta$ profile for the best fit model is shown as the dashed red curve. {\it Right:} At the best fit parameters (with $q \approx 0.66$), we plot equipotential contours. The stream track is also overplotted, color-coded by $\gamma$. The best fit equipotential contours are oblate, with a major axis that aligns closely with the disk. }
    \label{fig: NGC5907_PotentialConstraints}
\end{figure*} 

In this section we present preliminary constraints on the gravitational potential of NGC 5907 for both a single component potential model in \S\ref{sec: single_comp} and a two component potential model in \S\ref{sec: two_comp}. We utilize the Dragonfly data released with the \cite{2019ApJ...883L..32V} paper, reproduced in the top panel of Fig.~\ref{fig: ngc5907}. Red points indicate the most visually obvious segment of the stream, while cyan points indicate a tentative segment. We fit the red and cyan points using a differentiable spline, from which the unit curvature vector $\hat{\boldsymbol{\kappa}}$ along the stream can be estimated. The spline-based track and unit curvature vectors are shown in the bottom panel of Fig.~\ref{fig: ngc5907}. The green segment represents a region where the estimated track satisfies the zero curvature condition, namely, $\vert d\phi/d\ell \vert < \epsilon$ (\S\ref{sec: zero_curve} and \S\ref{sec: likelihood}). We treat the disk-halo angle, $\alpha$, as a free parameter in \S\ref{sec: two_comp}.  We present constraints on the full track (both the detection and tentative segment), and discuss where along the stream our constraints originate from.

\subsection{Single Component Potential}\label{sec: single_comp}
We adopt the logarithmic halo potential in Eq.~\ref{eq: log_pot} with $q_1 = 1$ and $q_2$ as a free parameter (simply referred to as $q$). The core radius is fixed to $r_c = 26~\rm{kpc}$, the same as in \citet{2019ApJ...883L..32V}. We also fit for the flattening direction, $\alpha$ (illustrated in Fig.~\ref{fig: ngc5907}, bottom panel). Because NGC 5907 is viewed nearly edge-on ($i \approx 87^\circ$; \citealt{1994Natur.370..441S}), we assume that the halo is viewed along a principal axis, with the flattening direction perpendicular to the line of sight, though possibly rotated with respect to the edge-on disk. We sample $q \in [0.4,2.5]$, and allow the flattening direction, $\alpha$, to vary between $-45^\circ$ and $45^\circ$. While $q < 1/\sqrt{2} \approx 0.7$ can produce negative mass densities near the origin for the logarithmic potential, the 3D geometry of the halo can bias the best fit $q$ towards lower values if the halo is oblate and not viewed along a principal axis. Furthermore, $ q < 1/\sqrt{2}$ can still provide a useful description of dark matter halos far from the origin ($\sim 10$s of kpc), where a large portion of the stream resides.

The likelihood surface in the space of rotation angles, $\alpha$, versus flattening, $q$, is shown in the left panel of Fig.~\ref{fig: NGC5907_PotentialConstraints}. The best fit (maximum likelihood estimate; MLE) parameters occur at the red point. A rotation angle of 0 occurs when the halo flattening axis is perpendicular to the disk. The best-fit rotation angle is $\alpha \sim -7~\rm{deg}$, though there is a large range in $\alpha$ that may also produce the stream's curvature. If the curvature vectors along the stream are mostly due to the disk's potential, then this is the expected behavior for a single component potential model (i.e., zero misalignment).

At the best-fit rotation angle, we plot the stream-acceleration angle, $\theta$, in the middle panel of Fig.~\ref{fig: NGC5907_PotentialConstraints} as a function of position along the stream track, $\gamma$. The $\theta$ versus $\gamma$ curves are color-coded by trial flattening parameters, $q$. For reference, the stream track is color-coded by $\gamma$ in the right panel. From the middle panel, we can determine where along the stream the constraints originate from. Namely, the regions of the stream closest to the disk (in projection) and the intermediate regions with $\gamma \approx 0$ rule out $q \lesssim 0.5$. The extended regions of the stream towards $\gamma \approx 0.25$ rule out $q \gtrsim 1.7$. Additional constraints come from the segment of the stream close to the disk (in projection), where we find that the tangent condition $\vert d\phi/d\ell \vert < \epsilon$ is satisfied. This indicates that planar accelerations pointing along the stream track should be preferred in this region  (\S\ref{sec: zero_curve}). The best fit potential (dashed red curve) is shown to simultaneously satisfy these constraints across the stream's length.

We overplot equipotential contours of the best-fit model on the image of NGC 5907 in Fig.~\ref{fig: NGC5907_PotentialConstraints}, right panel. We emphasize that our analysis knows nothing about the existence of the disk in photometry: the only input is the track of the stream, and the orientation of the potential model is allowed to vary along with its flattening. At the best fit parameters, equipotential contours align closely with the projected orientation of the disk. We find that our fits to a single-component potential model prefer a strongly oblate potential ($q \approx 0.6-0.8$), indicating a possibly substantial influence of the disk's potential on the orbit of the tidal stream. This is further supported by the small relative angle between the equipotential contours and the disk. 

There is an additional class of high-likelihood models in the left panel of Fig.~\ref{fig: NGC5907_PotentialConstraints} at more prolate $q \gtrsim 1.2$ with a rotation angle $\alpha \approx 45$. We have verified that when sampling $\alpha > 45~\rm{deg}$ that this mode continues until equipotential contours align with the disk. We take this to indicate a degeneracy in the likelihood surface, where an oblate potential can be expressed as a rotated prolate potential. We find similar behavior in \S\ref{sec: disk_halo_misalignment} on simulated data. Either mode suggests that equipotential contours are quite flattened when measured along an axis perpendicular to the disk. In the following section we explore a two-component potential model, with separate disk and halo components.

\subsection{Two Component Potential}\label{sec: two_comp}
In this section we follow the same approach presented in \S\ref{sec: disk_halo_misalignment} on simulated data, and attempt to constrain the parameters of a two-component potential model using the stream track of NGC 5907. We work in the same parameter space as in \S\ref{sec: disk_halo_misalignment}. 
Following  \citet{2008ApJ...689..184M} and \citet{2019ApJ...883L..32V}, we represent the disk with a Miyamoto-Nagai potential \citep{1975PASJ...27..533M} with scale-length 6.24~\rm{kpc} and scale-height 0.26~\rm{kpc}. The disk potential is rotated so that it is viewed edge-on and closely aligned with the major axis of the disk in observations, traced by its stellar mass  (Fig.~\ref{fig: ngc5907}). For the halo, we use the logarithmic potential (Eq.~\ref{eq: log_pot}) with scale-radius $r_c = 26~\rm{kpc}$ (following \citealt{2019ApJ...883L..32V}), $q_1 =1 $, and $q_2 = q$ (a free parameter). The halo potential is therefore axisymmetric. In what follows, we constrain the 2D orientation of the flattening axis (the disk-halo angle; $\alpha$), the flattening ($q$), and the mass ratio $\log_{10}(M_{\rm Disk}/M_{\rm Halo})$, where $M_{\rm Disk}/M_{\rm Halo}$ is the mass enclosed ratio within a 65~kpc radius, since this is roughly the extent of the stream in the $x-y$ plane. We again assume that the halo is viewed along a principal axis, so that the flattening axis is perpendicular to the line-of-sight. It is possible to relax these assumptions, but we leave this to future work.

\begin{figure}
    \centering
     \includegraphics[scale=0.42]{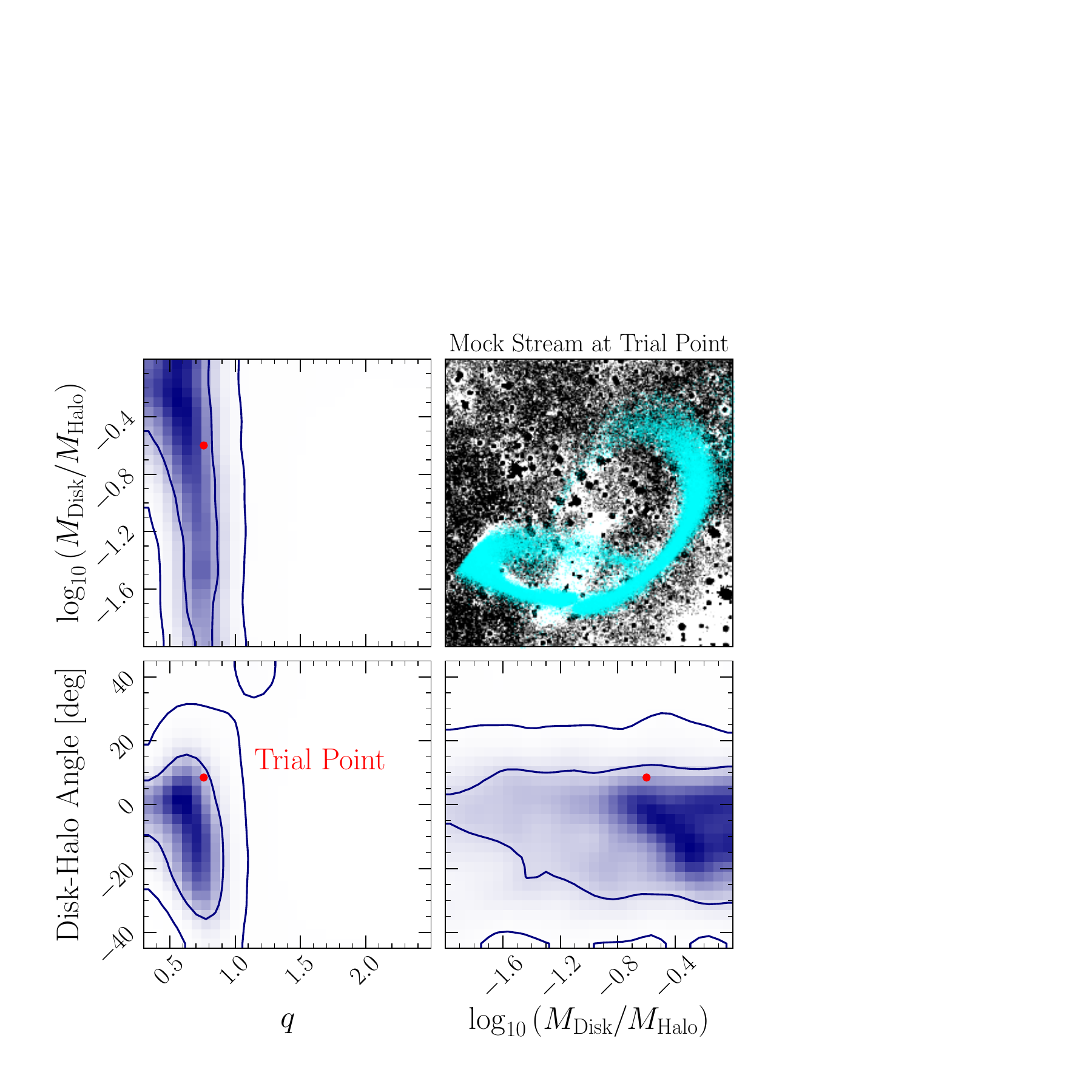}
    \caption{Constraints derived from the track of the stream surrounding NGC 5907 for a two-component potential model (disk + halo). The flattening ($q$), flattening axis (disk-halo angle), and mass ratio between the two potential components are all free parameters. Dark and light blue contours correspond to regions of 68 and 95\% confidence. We again find a preference for an oblate halo component, with $q \approx 0.8$. At lower disk-to-halo masses, the flattening parameter prefers slightly higher values. In the top right panel, we generate a mock stream in the two-component potential with parameters set to the trial point (red).  
    The morphology of the mock stream generated at the red point is well-matched to the stream surrounding NGC 5907. Both tentative and detection segments are captured by the mock stream with an oblate halo potential. }
    \label{fig: NGC5907_2CompModel}
\end{figure}

Posterior samples are obtained using the likelihood discussed in \S\ref{sec: likelihood} and the dynamic nested sampling package, \texttt{dynesty} \citep{2020MNRAS.493.3132S}. Constraints for the two-component potential are illustrated in Fig.~\ref{fig: NGC5907_2CompModel}, where contours correspond to regions of 68 and 95\% confidence. The mock stream in the top right panel is discussed in \S\ref{sec: validate}. When incorporating the disk we find a preference for a slightly higher $q = 0.76^{+0.10}_{-0.13}$ at 68\% confidence, when conditioning on $\log_{10}\left(M_{\rm Disk}/M_{\rm Halo}\right) < -1$ (reasonably, we expect the halo to have a dominant mass over the disk by at least an order of magnitude). Our fits are consistent with $q \approx 0.85$ at the $1\sigma$ level, corresponding to $q_{\rm density } \approx 0.55$. 

The mass ratio constraints neither confirm nor rule-out a substantial contribution from the dark matter halo surrounding NGC 5907. This is, in part, due to degeneracies induced by sampling over a wide range in stream line-of-sight distances (within $\pm 150~\rm{kpc}$ in this case). At large distances from the disk, the scale mass of the halo does not need to be dominant to reproduce the curvature of the stream, since the contribution of the disk to the acceleration field diminishes with distance. 

Still, it is interesting to consider the shape of the dark matter halo as a function of realistic disk-to-halo mass ratios. For instance, the top-left panel of Fig.~\ref{fig: NGC5907_2CompModel} implies an oblate halo component with $q \approx 0.85$ for $\log_{10}\left(M_{\rm disk} / M_{\rm Halo}\right) \approx -1$. The bottom right panel implies that there is not a significant misalignment between the halo and disk component. Both constraints indicate a preference for an oblate halo component, with a flattening direction within $\pm \sim 20^\circ$ of an axis orthogonal to the disk's midplane. These results do not necessarily imply that we are detecting the dark matter halo of NGC 5907. Instead, it could also be the case that the stream's curvature cannot rule out an oblate halo potential. Our analysis is consistent with both explanations. 

On the other hand, our constraints disfavor $q > 1$ at the $2\sigma$ level within the disk-halo angle interval considered. This indicates that we do not expect the halo component of the potential to be significantly different from the disk, at least interior to the stream. This is similar behavior to Milky Way potential constraints presented in \citet{2010ApJ...712..260K}, derived from the 6D phase-space distribution of the GD-1 stellar stream. In their work, the flattening of the halo model is attributed mostly to the influence of the disk. As we will discuss in \S\ref{sec: future}, we anticipate that kinematic data (e.g.,  radial velocity measurements) from the steam surrounding NGC 5907 will help disentangle the influence of the disk from that of the halo, and break degeneracies in our interpretation of Fig.~\ref{fig: NGC5907_2CompModel}.

\subsection{Validation with a Forward Model}\label{sec: validate}
We now use a forward model to generate a simulated stream in a potential consistent with our constraints in Fig.~\ref{fig: NGC5907_2CompModel}. We compare our results to \citet{2019ApJ...883L..32V}, who found that a simulated stream generated in a mildly prolate halo ($q=1.1$) reproduces the most prominent parts of the stream in observations. However, their model fails to reproduce the full extent of the stream ($\gamma\gtrsim 0.25$). In contrast, our analysis, which fits the entire observable length of the stream, prefers an oblate halo.

To validate that an oblate halo can produce the tidal feature surrounding NGC 5907, we select a trial point contained within the 68\% credible interval (red point in Fig.~\ref{fig: NGC5907_2CompModel}) and generate a mock stream at the trial point. The trial point is $\left(q,\log_{10}(M_{\rm Disk}/M_{\rm Halo}), \alpha\right) \approx \left(0.77,-0.6, 8.5^\circ\right)$. In order to generate mock streams, we calibrate the mass of the disk to yield a combined (disk plus halo) maximum circular velocity consistent with that used in \cite{2019ApJ...883L..32V}, derived from \citet{2019A&A...626A..56P}. For the disk-halo mass ratio considered here, this corresponds to $M_{\rm Disk} \approx 1.5\times 10^{11}M_{\rm Sun}$. The resulting halo mass at the trial point is then $M_{\rm Halo} \approx 10^{0.6}M_{\rm Disk} = 6\times10^{11}~M_{\rm Sun}$. For comparison, this corresponds to $M_{200} \approx 3\times10^{12}M_{\rm Sun}$ for the total potential, where $M_{200}$ is defined as the mass within a radius enclosing 200 times the critical density of the universe. This is similar, by construction, to rotation curve measurements from \citet{2019A&A...626A..56P}, who found $M_{200} \approx 10^{12} M_{\rm Sun}$.

We do not claim that these are in fact the absolute masses of the disk and halo components. Instead, we aim to show that the stream curvature is consistent with such absolute masses, based on our constraint of the mass ratio $M_{\rm Disk}/M_{\rm Halo}$. A larger mass difference is also supported by our constraints, e.g., with $\log_{10}\left(M_{\rm Disk}/ M_{\rm Halo}\right) = -1$. The same $M_{200}$ can be achieved with this mass ratio, by calibrating the absolute masses of each component (e.g., with the \citealt{2019A&A...626A..56P} measurements).

Using the trial point and the absolute masses for the two-component potential previously discussed, we generate a mock stream using the particle-spray method implemented in \texttt{Gala} \citep{gala}. We sample over a coarse grid in initial phase-space conditions, until a qualitative fit between the observed stream and the generated stream is achieved. The integration time is $\sim 6~\rm{Gyr}$ and the progenitor mass is taken to be $\sim 5\times10^8M_{\rm Sun}$. 

A mock stream generated using this approach is shown in the top right panel in Fig.~\ref{fig: NGC5907_2CompModel}. We find a strong qualitative agreement between the morphology of the stream imaged with the Dragonfly telephoto array and the mock stream generated in this section. The mock stream appears to have a similar tentative loop to the one shown in Fig.~\ref{fig: ngc5907}. This tentative loop has not been previously reproduced in the literature, though an oblate halo gives rise to this feature.

\section{Discussion}\label{sec: discussion}
We have shown that the curvature of extragalactic tidal streams can be used to constrain properties of the host potential. In this section we discuss our results, starting with the limitations of our analysis in \S\ref{sec: caveats}. We then consider which geometrical aspects of extragalactic tidal features will be most useful for constraining the potential of external galaxies in \S\ref{sec: what_makes_informative}. We draw comparisons to forward modeling techniques in \S\ref{sec: forward_model}, and discuss future directions in \S\ref{sec: future}.

\subsection{Caveats}\label{sec: caveats}
We first consider the stream track fitting routine discussed in \S\ref{sec: gen_Nbody} and Appendix~\ref{sec: fitting_routine}. The fitting routine currently implemented is not unique, nor optimal. For a given stream, there could be several tracks with slightly different curvature vectors that are compatible with photometry. One solution is to sample a posterior distribution over tracks, and propagate errors through to the model parameters of the potential. In \citet{Nibauer} an unsupervised stream track fitting algorithm is developed, though the algorithm utilizes velocity information that we are unlikely to measure in external galaxies. \cite{2022arXiv221200949S} developed another approach for fitting stream tracks, utilizing self-organizing maps and Kalman Filters. Both methods operate in the context of the Milky Way, and we leave the task of extending them to photometry of external galaxies to future work.

We now discuss whether a reliable distance to the host galaxy is required to carry out our analysis. The curvature-acceleration connection discussed in \S\ref{sec: modeling} is independent of distance to the host galaxy. However, a potential model is adopted to interpret the curvature of a stream. As long as the total potential model depends on distance to the host galaxy only through a scalar multiple (e.g., $\Phi \xrightarrow{} a\Phi$ when scaling distance), then unit vector accelerations in the plane of the sky are scale-free. This is the case for the Miyamoto-Nagai potential and logarithmic halo potential considered in this work. For a composite potential consisting of both a disk and halo component, the relative scale between characteristic lengths in each potential model can still influence the direction of unit vector accelerations in the plane of the sky (e.g., the scale length of the disk versus the core radius of the halo). The possibility of constraining the ratios of these parameters is left to future work. 

Throughout this work we make the assumption that a coherence condition is satisfied along the stream track (\S\ref{sec: curvature_accelerations}), allowing us to connect stream curvature to planar accelerations. In the Milky Way the gravitational potential is evolving, and this can influence the kinematics and morphology of tidal debris (e.g., \citealt{2019MNRAS.487.2685E,2021ApJ...923..149S,2021MNRAS.501.2279V,2023MNRAS.518..774L}). It is possible that the coherence condition can be satisfied even in time-dependent potentials, since we do not assume that proper motions along the stream are aligned with its track. Future work will be needed to determine the effect of a time-dependent potential on projected stream morphology. 

Throughout this work we have focused on stellar streams, though stellar shells have also been observed around external galaxies  (e.g., \citealt{1988ApJ...328...88S,2008ApJ...689..184M,2015MNRAS.446..120D}). Our method for constraining the potential of external galaxies can be extended to the case of shells, where the curvature of the projected shell boundary places limits on the direction of planar accelerations using similar arguments to those developed in \S\ref{sec: modeling}. We defer an exploration of this extension to future work.

\subsection{Sensitivity of Analysis to Stream Morphology: What features are most informative?}\label{sec: what_makes_informative}
\begin{figure*}[htp!]
    \centering
    \includegraphics[scale=0.72]{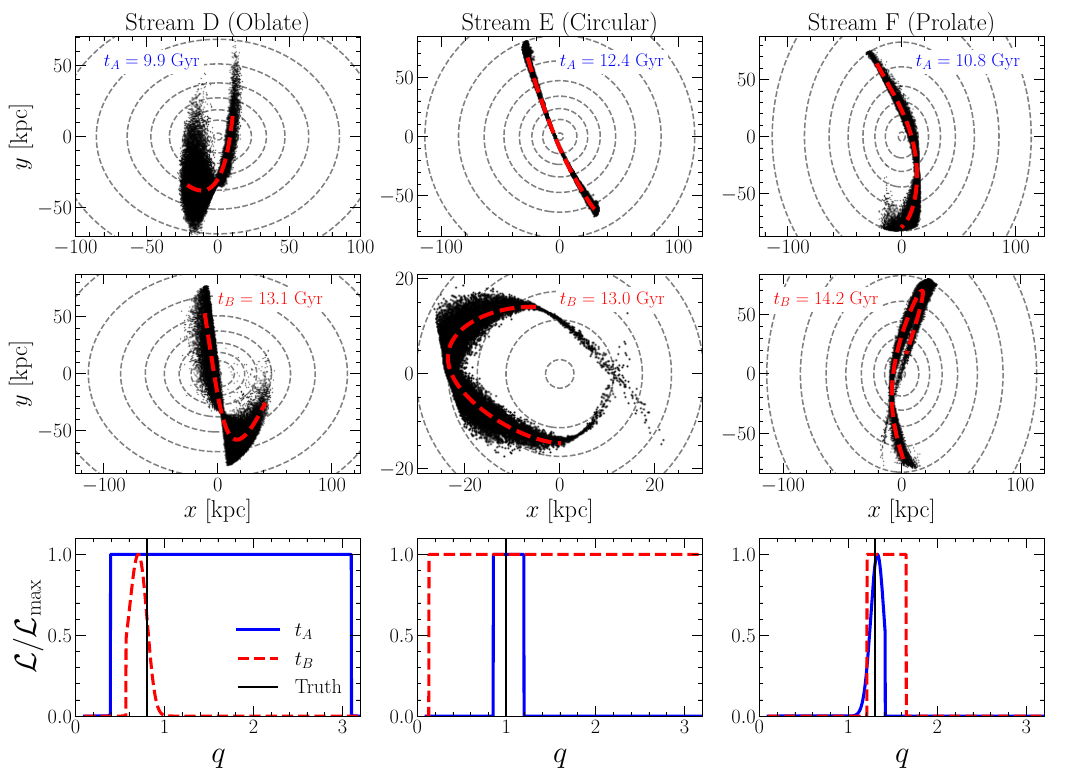}
    \caption{Potential constraining power as a function of orbital phase. Each column corresponds to the same $N-$body stream (D, E, F), though at two different snapshots: $t_A$ (blue) and $t_B$ (red). Dashed red curves indicate the stream track for each panel, and dashed gray curves are equipotential contours. The curvature of the tracks is used to estimated the $y-$axis flattening parameter, $q$. The likelihood function for each stream is shown in the bottom row, where blue corresponds to the $t_A$ snapshot, and red corresponds to the $t_B$ snapshot. The true flattening parameter is shown as the vertical black line. The ability to constrain the potential from only the morphology of a tidal stream is strongly dependent on when the tidal feature is observed throughout its evolution.}
    \label{fig: vary_snaps}
\end{figure*} 

In this section we discuss what morphological properties of tidal streams will be most useful for constraining the potential of external galaxies, based on our analytic model in \S\ref{sec: modeling} and the tests presented throughout the paper. For instance, in Fig.~\ref{fig: 3Stream_Constraints}, stream B seems to suggest that our method has poor constraining power for spherical potentials since the likelihood function supports a wide range in flattening parameters. Alternatively, it could also be the case that the constraining power of our analysis is tied to the specific geometrical properties of the tidal feature considered, and therefore when it is observed throughout its evolution. We now carry out an exercise to test how constraints on the potential vary as a function of stream morphology, by analyzing the same stream at two different stages in its evolution. 

In Fig.~\ref{fig: vary_snaps} we test how the observed morphology of a stream affects our ability to place informative constraints on the underlying potential. In each column we plot the same stream, though ``observed" at two different epochs indicated by times $t_A$ and $t_B$. The particular snapshots are chosen to demonstrate how the likelihood function can change significantly as a function of different stream lengths and geometries. The background potential varies across streams D-F, though for a given column the potential model is exactly the same. 

For each snapshot, the spline-based stream-track is shown by the red dashed-curve. We plot the corresponding likelihood functions for each system in the bottom row. We utilize the full likelihood introduced in \S\ref{sec: likelihood}, with the tangent condition implemented when the stream passes through a region without curvature. The blue likelihood curve corresponds to snapshot $t_A$, while the dashed red curve corresponds to snapshot $t_B$. The true flattening parameter for each case is shown as the black vertical line.

For the same stream observed at two different epochs, the width of the likelihood function can vary significantly. This is most evident for Streams D and E in Fig.~\ref{fig: vary_snaps}, which display a significant change in morphology between snapshots $t_A$ and $t_B$. While all snapshots yield constraints that are consistent with the true halo shape, streams with segments of negligible curvature, inflection points, or abrupt apocentric passages tend to provide 
much stronger constraints on halo shape parameters. This aligns with the analytic treatment introduced in \S\ref{sec: modeling}: streams with tangent vectors that change direction quickly along the track sample the largest range of curvature vectors in the plane. The range of compatible accelerations for a given potential will be severely limited, since the constraint $\vert \theta \vert < \pi/2$ becomes increasingly restrictive if the stream samples several directions over a small area. This is typically the case for streams with sharp apocenters, as for streams D and F at their respective $t_B$. 

Our analysis is also powerful when a segment of the stream has negligible curvature (e.g., the tidal feature surrounding Centaurus A forward modeled in \citealt{2022ApJ...941...19P}). In this case, we employ the tangent condition discussed in \S\ref{sec: zero_curve}, which constrains the planar acceleration vector to point along the local, curvature-free segment of the tidal feature. Our analysis is most informative when a stream has segments with negligible curvature, and with significant curvature. This occurs for stream D at time $t_B$ in Fig.~\ref{fig: vary_snaps}. Towards the left end of the track, the curvature becomes negligible (i.e., $\vert d\phi/d\ell\vert<\epsilon$) and the tangent condition is applied. The intermediate region of the track has a sharp turn, so the tangent vector changes rapidly over a small spatial scale. The result is a tight constraint on the shape of the gravitational potential, since collectively, there are only a small range of continuous acceleration fields that can satisfy these constraints simultaneously across a single stream. Alternatively, multiple streams surrounding one galaxy with different constraining features can be used to fit the potential through a joint likelihood (\S\ref{sec: triaxial_potential}). 

Finally, we highlight a third regime that is illustrated in Fig.~\ref{fig: vary_snaps}, stream E at $t_A$. The snapshot is taken at a moment when the stream passes near the origin at an angle inclined from the $x$-axis. In this snapshot, curvature vectors are pointing up and to the right. For a potential that is strongly oblate, the bottom right segment of the stream curves in the correct direction (towards the central mass density), while the top left segment curves away from the central mass density. In this case, the top left segment of the stream is incompatible with an oblate potential. For a strongly prolate potential, the converse is true: the top left segment of the stream curves towards the central mass density, while the bottom right segment curves away. Therefore, the bottom right segment of the stream is incompatible with a strongly prolate potential. From the likelihood plot in the bottom row, this configuration is enough to rule out either extreme (i.e., an oblate or prolate potential), and is best satisfied by $q \approx 1$; in between prolate and oblate. Thus, extended streams passing near the central mass density  of the host (in projection) provide another case where the potential is well constrained.

Collectively, the above scenarios can occur over a wide range of potentials. The constraining power of our analysis is tied to the presence of morphological features along a stream, which is predominantly determined by the stream's length and our viewing angles, rather than the halo shape. As long as the constraining morphologies discussed above are present, we can, in principle, constrain even complicated potential models with many shape parameters.

\subsection{Comparison to Forward Modeling}\label{sec: forward_model}
In this section we discuss how our method compares to forward modeling techniques for stream formation in a specified potential. 

A common method for forward modeling streams is the ``particle spray" technique \citep{2015MNRAS.452..301F}, where test particles are released from a parent orbit near the Lagrange points of a progenitor with a specified mass. Other forward models adopt similar techniques, though with different stream formation prescriptions (e.g., \citealt{1999ApJ...512L.109J,2006MNRAS.366.1012F,2010ApJ...712..260K,2014ApJ...794....4P,2012MNRAS.420.2700K,2014ApJ...795...94B,2014MNRAS.445.3788G}). 

Forward models simulate the full time-evolution of the released particles, and can reproduce a wide range of stream morphologies. Additionally, these techniques can utilize the observed width of the stream to constrain the potential, along with its surface brightness profile (and therefore number density). The effects of a time-dependent potential can also be incorporated, along with perturbations from substructure (e.g., dark matter subhalos; \citealt{2019ApJ...880...38B}).

On the other hand, forward models have a large number of nuisance parameters that specify the full 6D phase-space initial conditions of the progenitor, progenitor mass, mass loss rates, integration times, the orientation of the system with respect to the line of sight, and the parameters of the adopted potential model. If the object of interest is the gravitational potential of the host galaxy, attempting to infer all of these quantities simultaneously from only projected stream morphology introduces strong degeneracies with the parameters that we wish to constrain. These degeneracies are only lifted with the inclusion of kinematic data. In the absence of this information, it is likely that the number of modeling degrees of freedom are higher than what the data can realistically constrain (e.g., the same stream morphology can be achieved for different halo masses; \citealt{2022ApJ...941...19P}). The method also scales poorly with the number of streams, and does not indicate which parts of the stream(s) considered are driving constraints on the potential.

In contrast, the method presented in this work does not model initial conditions, nor the unobserved velocity components along the stream. The only required input is the shape of the current stream track, and an estimate of the track's curvature along the stream. We do not rely on orbit integrations, stream formation prescriptions, nor prior knowledge of the progenitor's mass or phase-space position. Instead, the computational cost is front loaded to the track-fitting routine. Once a track is fit to the stream(s) of interest, varying the parameters of a potential model and the orientation of the potential is inexpensive, since we only compare the direction of acceleration vectors to the observed stream curvature. Furthermore, our approach also indicates where along the stream constraints on the potential originate from. Our inference can be performed simultaneously for multiple streams around a single galaxy (e.g., \citealt{2016ApJ...823...19C}, and some of the candidate mergers in \citealt{2023arXiv230204471G}), or for many galaxies with streams (e.g., \citealt{1998ApJ...504L..23S,2001Natur.412...49I,2009ApJ...692..955M,2016ApJ...823...19C, 2018A&A...614A.143M, 2018ApJ...866..103K,2022A&A...662A.124S,2023arXiv230204471G}). This gives rise to the possibility of constraining dark matter halos at the population level, as we will discuss in \S\ref{sec: future}. 

A distinct advantage of our approach over forward modeling is that our model in \S\ref{sec: modeling} is very general, though still provides meaningful insights on what can actually be inferred about the potential from extragalactic stream morphology. For instance, the curvature-acceleration connection derived in \S\ref{sec: curvature_accelerations} tells us that from stream curvature, we can only constrain properties of the potential that influence the direction of the acceleration field when projected onto the plane of the sky. In this case, it is clear that the absolute mass of a galaxy and its halo cannot be determined (without imposing priors on the progenitor's initial conditions), but a mass ratio between the two components can be constrained. Shape properties such as the flattening direction, and even 3D orientation of the halo or disk can also alter the direction of planar acceleration vectors making these good parameters to target. We can also predict which geometrical aspect of tidal features will be most useful for constraining halo shape parameters, such as inflection points or segments of a stream that appear linear in projection.

There are limitations to our method as well. For example, we do not constrain the initial conditions of the progenitor, nor its present-day location. However, if the only available measurements are of the stream's projected morphology, we do not expect that initial conditions can be determined with high confidence. Our approach does not impose priors on the velocity of the progenitor, or stars released from the progenitor. This means we are more agnostic to the progenitor mass, though stream-width could encode information about this quantity (e.g.,  \citealt{2016MNRAS.461.1590E}). We have also neglected modeling the stream's surface brightness profile, since we have only considered stream curvature. Incorporating surface brightness information is possible in our analysis, though we defer this to future work. Finally, we have considered the realistic scenario of having no kinematic information of the stream but only its projected morphology. For a limited number of extragalactic streams, radial velocities or distance information has been measured (e.g., from the progenitor globular cluster; \citealt{2014MNRAS.442.2929V}; surface brightness fluctuations; \citealt{2016ApJ...824...35T}; and tip of the red-giant branch measurements; \citealt{2019ApJ...872...80C}). When available, these measurements can be incorporated in our constraints as discussed in \S\ref{sec: future}. However, the present work considers the more ubiquitous scenario of having only measured the stream morphology from imaging.

\subsection{Future Directions}\label{sec: future}
The fundamental properties of dark matter halos are their mass and shape; while there are many techniques to measure halo masses 
(e.g., \citealt{1997ApJ...488..702R,2009MNRAS.393..329N,2010MNRAS.406..264W,2019A&A...626A..56P}), measurements of halo shapes beyond spherical models have been more limited.
In the extragalactic context, weak lensing has been the primary method for constraining halo shapes by stacking galaxy images and measuring the resulting shear field (e.g., \citealt{2000ApJ...538L.113N,2006MNRAS.370.1008M}). Extragalactic tidal features provide a new method to measure halo shape parameters at the level of individual galaxies, and across an ensemble.

\citet{2022ApJ...941...19P} showed that with only a single radial velocity measured from an extragalactic stream (in addition to stream morphology), the halo mass is much better constrained. While the analysis presented in this work is entirely concerned with stream curvature and its connection to halo shape parameters, radial velocity information can also be included in our approach. By calibrating the rotation curve of our morphology-based model against a radial velocity measurement from the stream, there is a clear path forward to self-consistently incorporate kinematic information in our constraints. Indeed, we have already carried out a similar analysis in \S\ref{sec: application_ngc59076}, where we use previous measurements of the rotation curve from NGC 5907 \citep{2019A&A...626A..56P} to calibrate the inferred mass ratio $M_{\rm Disk}/ M_{\rm Halo}$. If a radial velocity is supplied to our model at the stream's radius, we anticipate that this information could also be utilized to better inform halo shape parameters, and help disentangle the effect of the halo from the disk. We leave this to future work.

Perhaps the most exciting future direction to consider is a population level description of dark matter halo shapes (e.g., axis ratios and orientations), probed by the morphology of extragalactic streams. From the 1000s of extragalactic tidal features expected to be observed over the next decade with The Rubin Observatory \citep{2019ApJ...873..111I}, {\it Euclid} \citep{2016SPIE.9904E..0OR}, and {\it Roman} \citep{2013arXiv1305.5422S}, constraints on dark matter halo shape parameters can be combined at the population level to obtain a global view of dark matter halo properties. While stream morphology based constraints on the dark matter halos of individual galaxies can be weak or degenerate with viewing angle, a population level (e.g., hierarchical) model that combines information across many galaxies and streams will be better constrained. Our analysis is suited for this purpose, since all that is required is the track of a stream and its curvature vectors. From a population level description of dark matter halo morphologies, comparisons can be drawn to cosmological simulations which make important predictions for dark matter halos shapes in $\Lambda\rm{CDM}$  (e.g., \citealt{2006MNRAS.367.1781A,2012JCAP...05..030S, 2019MNRAS.484..476C, 2022arXiv221208880P}).

We have worked under the assumption of cold dark matter, though our analysis could be sensitive to other descriptions that alter the acceleration profile of a galaxy (e.g., Modified Newtonian Dynamics; \citealt{1983ApJ...270..365M}). We leave an exploration of extragalactic streams and alternative descriptions of the gravitational acceleration field to future work.

\section{Summary and Conclusion}\label{sec: conclude}
We have drawn an analytic connection between the curvature of tidal streams and the underlying acceleration field of the host galaxy. The curvature-acceleration connection can be used to identify gravitational potential models that are consistent  with the projected 2D morphology of the tidal feature considered. We have shown that in order to connect stream curvature to accelerations---and therefore the potential---the stream does not need to trace a single stellar orbit. In \S\ref{sec: likelihood} we devised a likelihood to identify the range of potential models that are compatible with the local curvature direction(s) of a stream. The only required input to our analysis is a differentiable stream track, from which curvature vectors can be estimated. The method is computationally inexpensive, and does not rely on forward modeling techniques. Because of this, our analysis scales well with the number of streams considered.

We have demonstrated the method on stellar streams generated in $N-$body simulations with a static background potential. From 2D spatial snapshots of these simulations, we recover flattening parameters of the host halo, disk-halo misalignment angles, and in some cases mass ratios between different potential components. The best constrained parameter is the halo flattening, while constraints on misalignment angles and mass ratios are weaker. We have also shown that in some cases the 3D orientation of the halo can be constrained, and that limits can also be placed on the axis ratios of a triaxial halo model from only projected stream morphology. These constraints have significant degeneracies, though still allow us to rule out a range of possible geometries. Modeling multiple streams with constraining features can help place more stringent constraints on the 3D potential.

Our analytic treatment implies that streams with a change in concavity or a linear segment are most informative, since in these regions planar accelerations align with the stream track. Streams with tightly wound loops are also informative, since their curvature vectors sample many directions over a small area in the sky. 

We apply our analysis to the stellar stream surrounding the edge-on galaxy NGC 5907, and find a preference for an oblate halo potential ($q = 0.76^{+0.10}_{-0.13}$ when conditioning on $M_{\rm Disk} / M_{\rm Halo} < 0.1$). Mock streams generated in potentials consistent with our constraints successfully recover the observed stream morphology in full.

There are already several observations of extragalactic streams (e.g., \citealt{1998ApJ...504L..23S,2001Natur.412...49I,2009ApJ...692..955M,2016ApJ...823...19C, 2018A&A...614A.143M, 2018ApJ...866..103K,2023arXiv230204471G}), with 1000s more expected from The Rubin Observatory \citep{2019ApJ...873..111I}, {\it Roman} \citep{2013arXiv1305.5422S}, and {\it Euclid} \citep{2011arXiv1110.3193L}. Traditional methods for constraining the potential using extragalactic streams will be hampered by the lack of kinematic information and distance tracks, since we will often only have measured the projected stream morphology from imaging. The approach developed in this work is well-suited to this limited data scenario, and is computationally inexpensive. The method can be applied to a large number of galaxies with tidal streams, providing the means to measure the 3D shapes of dark matter halos at the population level. This enables future tests of $\Lambda\rm{CDM}$ using the morphology of extragalactic streams, by comparing the inferred distribution of dark matter halo geometries to theoretical predictions from cosmological simulations. 

\section*{Acknowledgements}
We are very grateful to Sarah Pearson for a careful read and discussion of the manuscript. We are also grateful to David Spergel, Vasily Belokurov, Adrian Price-Whelan, Shirley Ho, Jenny Greene, Tomer Yavetz, Jeremy Goodman, Shaunak Modak, and Maureen Iplenski for helpful discussions. JN would like to acknowledge the hospitality of the Carnegie Observatories, where a portion of this work was completed.

\software{Numpy \citep{ harris2020array}, Matplotlib \citep{Hunter:2007}, PyTorch \citep{NEURIPS2019_9015}, SciPy \citep{2020SciPy-NMeth}, Gala \citep{adrian_price_whelan_2020_4159870}, Astropy \citep{astropy:2018}, NEMO \citep{1995ASPC...77..398T}, Corner \citep{corner}. }

\begin{appendix}
\section{Stream Track Fitting Routine}\label{sec: fitting_routine}
The modeling approach introduced in \S\ref{sec: modeling} relies on estimates of unit curvature vectors along the projected stream track. The unit curvature vector is related to the second derivative of position along the projected track, as expressed in Eq.~\ref{eq: kappa_def}. We estimate these derivatives by fitting a smooth, differentiable spline to a series of representative points along the $N-$body stream in the plane of the sky. Our fitting routine works in two steps. For a given snapshot, we first bin particle positions in the $x-y$ plane and take the average $(x,y)$ coordinate in each bin. For the binned coordinates, we select nearby neighbors to construct a fiducial stream track, or automatically join neighboring bins based on the discussion in \S\ref{sec: fitting_stream_track}. This fiducial track is used to assign an ordering to the stars in the stream, by measuring the minimum distance between stream particles and the fiducial track. Each star is assigned an ordering parameter $\gamma \in [-1,1]$ based on where along the fiducial track they are closest to. The difference in $\gamma$-values for subsequent stars is set to a constant, such that the interval $\gamma \in [-1,1]$  is partitioned into $N-1$ equal segments (where $N$ is the number of stars belonging to the tidal feature). For instance, stars towards one end of the track will have $\gamma$ closer to $-1$, while stars located at the opposite end will have $\gamma$ close to $1$. The final track used in the analysis is determined by fitting a spline to the now ordered stream. We utilize the \texttt{Scipy} interpolation library to fit a smooth spline to the projected stream, and differentiate the track using a representation of the same spline in \texttt{PyTorch} \citep{NEURIPS2019_9015} with automatic differentiation enabled. We fit a smoothing spline of degree $k \in [3,4,5]$, with a large (e.g., $10^5-10^6$) smoothing factor to ensure that the projected track characterizes the average shape of the stream rather than passing exactly through the data points. 

This approach works for observations that resolve individual stars belonging to a tidal feature, however this is generally not the case. For unresolved streams, binned (i.e., average) photometric measurements can be utilized to fit a track to the tidal feature. As long as photometric measurements can capture the average shape of the stream, we expect our analysis to be applicable. This is the approach we take in \S\ref{sec: application_ngc59076}, where we analyze a real tidal feature surrounding the galaxy NGC 5907. In a future work, we will consider more rigorous and automated techniques for fitting flexible and differentiable curves to the observed population of tidal debris (e.g., \citealt{2022arXiv221200949S}).

\section{Hyperparameters}\label{app: Hyper_Params}
We now discuss the various hyperparameters used throughout this work, and their values. Most hyperparameters relate to the stream track and our fitting routine. We could consider a Bayesian determination of the stream track, allowing us to treat the hyperparameters in a probabilistic framework. However, the present work is aimed at introducing and validating the basic principles discussed in \S\ref{sec: modeling}, so we postpone a more thorough exploration of how to effectively fit the stream track to future work.

In \S\ref{sec: likelihood} we introduced the hyperparameters $\Delta D$, $\epsilon$, and $\sigma_{\theta_T}$. These correspond to the maximum change in line-of-sight distance between adjacent evaluation points along the stream track, the zero curvature threshold ($\vert d\phi/d\ell \vert < \epsilon$), and our prior uncertainty on the direction of the stream track when the tangent condition is satisfied, respectively. For $\Delta D$, we adopt $30~\rm{kpc}$, since this is a reasonable upper bound for the line-of-sight distance between adjacent stream segments in our $N-$body simulations. For a stream at $\approx 17~\rm{Mpc}$ (i.e, NGC 5907 considered in \S\ref{sec: application_ngc59076}), a $0.035~\rm{deg}$ change in the direction of the stream track per $\rm{kpc}$ corresponds to an $\epsilon \sim 10$. We adopt this value for $\epsilon$, and $\sigma_{\theta_T} = 10~\rm{deg}$. In our tests on simulated data, we find this value of $\epsilon$ successfully identifies regions of the stream track with negligible curvature. Our usage of a relatively large $\sigma_{\theta_T}$ ensures that the tangent condition will not strongly disfavor planar acceleration vectors with angular separations of $\sim 1-4~\rm{deg}$ from the local stream-track. 

\section{Expanded Likelihood}\label{app: expanded_Likelihood}
In this appendix we write out the full likelihood utilized in the analysis. The likelihood is already derived in \S\ref{sec: likelihood}, here we simply expand Eq.~\ref{eq: likelihood} to demonstrate a relationship with the trinomial distribution.  Each of the conditional probabilities in Eq.~\ref{eq: C12_likelihood} and Eq.~\ref{eq: NoCurve_likelihood} are zero if the condition ($C_j$) is not satisfied. That is, for each evaluation point along the stream (each $\hat{\boldsymbol{\kappa}}_i$) only one term in the summation of Eq.~\ref{eq: L_i} survives. In this way these probabilities have similar behavior to indicator variables, which are either zero or one. Eq.~\ref{eq: fractions} is the fraction of times each conditional probability is non-zero. This means the full likelihood (Eq.~\ref{eq: likelihood}) can be expanded as
\begin{multline}\label{eq: expanded_likelihood}
    \prod_{i=1}^N P\left(\hat{\boldsymbol{\kappa}}_i |\boldsymbol{m}, z(\gamma) \right) =  \\ \left[ f_{1} \right]^{N f_{1}} \left[ f_{2} \right]^{N f_{2}} \left[f_{3}\right]^{Nf_{3}} 
     \\ \times \prod_{\hat{\boldsymbol{\kappa}}~\mathrm{undefined}} \left[ \mathcal{N}\left(\theta_T |0, \sigma_{\theta_T} \right)
     \right],
\end{multline}
where $N$ is the total number of evaluation points along the stream track and the product is over all $N\times f_3$ evaluation points with undefined curvature vectors. The parameter $f_1$ is the fraction of evaluation points along the stream for which the curvature and planar acceleration vectors are consistent, $f_2$ is the fraction of evaluation points with inconsistent curvature and planar acceleration vectors, and $f_3$ is the fraction of evaluation points with no curvature. Note that $f_3$ is independent of the model parameters $\boldsymbol{m}$, since the number of evaluation points without curvature does not change with the choice of a potential model. The terms with $f_1, f_2, f_3$ in Eq.~\ref{eq: expanded_likelihood} collectively form the trinomial distribution, since the analysis has three possible binary outcomes:  either a trial acceleration is consistent with the curvature vector, not consistent, or the curvature vector is undefined. The last case introduces an additional probability component that is continuous, since in this scenario accelerations should align close to the stream track.

\section{Specifying Edge Cases in Distance Sampling}\label{app: likelihood_edge}
In \S\ref{sec: likelihood} we discuss the likelihood used in this work, which performs a maximization over the line-of-sight coordinate, $z$. In \S\ref{sec: distance_gradients} we discuss how the line-of-sight coordinate is sampled to ensure the possibility of a continuous distance distribution along a stream. In general, there will be some distance interval, $D_i$, over which the stream segment indexed $i$ has $ \vert \theta_i \vert < \pi/2$. This interval for the preceding segment, indexed $i-1$, must overlap within some threshold $\Delta D$ of $D_i$ in order to ensure the possibility of a continuous distance track with a well-behaved gradient. Therefore, the range of allowable distances for the $i^{\rm th}$ evaluation point is set by the preceding evaluation point with index $i-1$. We refer to the range of acceptable distances for the $i^{\rm th}$ evaluation point with the $\Delta D$ “padding” as the acceptable distance interval, represented by $\mathcal{A}_{i-1}$ (since the interval depends on the preceding segment).

We now discuss a few edge cases in this sampling scheme, and how they are addressed:
\begin{itemize}
    \item \textbf{The first evaluation point ($i = 0$):} For the first evaluation point the acceptable distance interval that we sample over is the full distance interval $z \in [z_{\rm min}, z_{\rm max}]$. We usually take this interval as the maximum extent of the stream in the plane of the sky. For NGC 5907 in \S\ref{sec: application_ngc59076}, we use a more conservative interval with $z \in [-150,150]~\rm{kpc}$.

    \item \textbf{Extended Stream Segment with Negligible Curvature:}
     If both evaluation points $i$ and $i-1$ have negligible curvature, we define 
\begin{equation}
    \hat{z}_i \equiv \underset{ z \in \mathcal{A}_{i-1}}{\mathrm{argmax}}\left[ \mathcal{N}(\theta_T| 0, \sigma_{\theta_T})\right],
\end{equation}
which is the most probable line-of-sight distance for the $i^{\rm th}$ evaluation point. The acceptable distance interval $\mathcal{A}_{i-1}$ is taken to be within $100~\rm{kpc}$ of the preceding evaluation point with  distance $\hat{z}_{i-1}$. The choice of $100~\rm{kpc}$ is conservative, and ensures that we sample over a range of possible distances to identify a potential model that is compatible with the stream track.

\end{itemize}

\end{appendix}

\bibliography{thebib}
\end{document}